\documentclass[12pt]{iopart}

\usepackage{amssymb,amsthm}
\usepackage[english]{babel}
\usepackage{graphicx}
\usepackage{color}
\usepackage{tabularx}
\usepackage{tikz}
\let\oldhat\hat
\renewcommand{\vec}[1]{\mathbf{#1}}
\renewcommand{\hat}[1]{\oldhat{\mathbf{#1}}}

\begin{document}

\title[Persistent polymer in lattice disorder]{Influence of lattice disorder on the structure of persistent polymer
chains}

\author{Sebastian Sch\"obl, Johannes Zierenberg and Wolfhard Janke}
\address{Institut f\"ur Theoretische Physik and Centre for Theoretical Sciences (NTZ), \\ 
Universit\"at Leipzig, Postfach 100920, D--04009 Leipzig, Germany}
\ead{sebastian.schoebl@itp.uni-leipzig.de}

\begin{abstract}

  We study the static properties of a semiflexible polymer exposed to a quenched random
  environment by means of computer simulations.  The polymer is modeled as two-dimensional
  Heisenberg chain. For the random environment we consider hard disks arranged on a square
  lattice.  We apply an off-lattice growth algorithm as well as the multicanonical Monte Carlo
  method to investigate the influence of both disorder occupation probability and polymer
  stiffness on the equilibrium properties of the polymer.  We show that the additional length
  scale induced by the stiffness of the polymer extends the well-known phenomenology
  considerably. The polymer's response to the disorder is either contraction or extension
  depending on the ratio of polymer stiffness and void space extension. Additionally, the
  periodic structure of the lattice is reflected in the observables that characterize the
  polymer.

\end{abstract}
\submitto{JPA}

\maketitle


\section{Introduction}

The conformational properties of polymers exposed to disordered media are
strongly affected by the surrounding disorder potential. For the case of
flexible polymers, the impact of disorder on polymers has already been widely
discussed [1--9].
The special case of geometrical constraining environments has been investigated
in e.g.\ \cite{Baumgaertner1987,Schoebl2011}.  It is expected that geometrical
restrictions to chain conformations also play a crucial role for biological
systems. In these systems, polymers may no longer be assumed flexible and models
of moderately stiff polymers, called {\em semiflexible} polymers, are
introduced.  The stiffness is characterized by the persistence length $l_p$. On
length scales shorter than the persistence length, the polymers behave like
stiff rods, on longer scales they exhibit entropic flexibility and random
coiling occurs. The geometrical restrictions of the environment along with the
intrinsic stiffness of the polymers lead to an interesting phenomenology, which,
in contrast to the case of flexible polymers, is much less understood for
semiflexible polymers [10--14].

In this work we examine the equilibrium properties of a pinned {\em
semiflexible} polymer exposed to a quenched random potential consisting of hard
disks.  The disks are arranged on the sites of a square lattice. We build up on
\cite{Schoebl2011}, where {\em flexible} polymers exposed to hard-disk disorder
assembled on the sites of a square lattice were investigated. We extend the
polymer model to comprise bending stiffness. The appropriate polymer model is the
Heisenberg chain model. Additionally, we consider the effect of leaving the
constraint of a fixed starting point.

The rest of the paper is organized as follows. In \sref{sec:modAndMeth} we
describe the polymer model and the assumed disorder configurations.
\Sref{sec:algorithm} is devoted to the employed simulation algorithms and in
\sref{sec:obsParamsTestCases} we define the measured observables, discuss the
simulation parameters, and present a few test cases. Our main results are
contained in \sref{sec:quenchedAverage}, where we first discuss the low
disorder-density case and then the more intricate high-density regime. We
conclude this section with a few remarks on the impact of the hard-disk diameter
and the initial pinpoint. Finally, in \sref{sec:conclusions} we summarize our
main findings.


\section{Model}\label{sec:modAndMeth}


\subsection{Polymer model}\label{sec:polymerModel}

Effectively, the Heisenberg chain is a bead-stick model consisting of $N+1$
beads at positions $\vec r_i$ connected by bonds of fixed length $b$. Therefore,
the contour has a length of $L=N b$. Our considerations are made for the case of
two dimensions and a phantom chain where self-avoiding constraints are
neglected. The connecting line between two monomers defines a unit tangent vector $\vec
t_i = (\vec r_{i+1}-\vec r_i)/b$.  The elastic properties are governed by the
bending energy
\begin{equation} \mathcal{H} = -J
  \sum_{i=1}^{N-1} \vec t_i \vec t_{i+1}, 
  \label{eq:freeChainHamiltonian}
\end{equation}
where $\vec t_i \vec t_{i+1}=\cos(\theta_{i,i+1})$ determines the angle between
neighboring bonds and $J>0$ is a coupling constant.  The correlations between
the two-dimensional tangent vectors of the free Heisenberg chain decay at
inverse temperature $\beta=1/k_{\mathrm B}T$ as \cite{Thompson}
\begin{equation}
  \langle \vec t_i \vec t_{i+k} \rangle = \left [\frac{I_1(\beta J)}{I_0(\beta J)} \right ]^k,
  \label{eq:ratioOfBesselFunctions}
\end{equation}
where $I_{\mu}(x)$ is the modified Bessel function of the first kind of order
$\mu$. 

Carrying out the continuum limit of the Heisenberg chain by taking
\eref{eq:freeChainHamiltonian} and letting $N,J\rightarrow \infty$ while
$b\rightarrow 0$ with $Jb=\mathrm{const}.$\ and $Nb=L$ (constant length
constraint), transfers \eref{eq:freeChainHamiltonian} up to a constant into 
\begin{equation}
  \mathcal{H}=\frac{\kappa}{2}\int_0^L ds \left
  (\frac{\partial^2 \vec R(s)}{\partial s^2} \right ) ^2
  \label{eq:WLCHamiltonianContinuumLimit}
\end{equation}
with $\kappa=Jb$ being the bending stiffness and $\vec R(s)$ describing the contour
parametrized by arc length $s$.  \Eref{eq:WLCHamiltonianContinuumLimit} is the
Hamiltonian of the worm-like chain, also called Kratky-Porod model
\cite{KratkyPorod}, one of the most famous and widely spread models for
treating semiflexible polymers analytically.

A central property of the worm-like chain is its {\it persistence
length} $l_p$, which is the tangent vector correlation length \cite{DoiEdwards}, 
\begin{equation}
  \langle \vec t(0) \vec t(s) \rangle = e^{-s/l_p},
  \label{eq:tangentCorrelationsGeneral}
\end{equation}
where $\vec t(s)=\partial \vec R(s)/\partial s$. In the continuum
limit of the Heisenberg chain Hamiltonian, we consider the following
approximation.  For large $\beta J$ or small $b$, and therefore large $N$, the
modified Bessel function in \eref{eq:ratioOfBesselFunctions} yields
\cite{Abramowitz}:
\begin{equation}
  I_{\mu}(x) \approx \frac{e^x}{\sqrt{2\pi x}} \left \{ 1-
  \frac{4\mu^2 -1}{8x} + \frac{(4\mu^2-1)(4\mu^2-9)}{2! 8x^2} -
  \mathcal{O}(x^{-3}) \right \}.
  \label{eq:BesselApproximation}
\end{equation}
Thus, for large $\beta J \propto N$ and $l=kb$ one finds for the tangent
correlations by inserting \eref{eq:BesselApproximation} into
\eref{eq:ratioOfBesselFunctions} to leading order:
\begin{equation}
  \langle \vec t_i \vec t_{i+k} \rangle = \mathrm{exp} \left
  (-\frac{k_BT}{2Jb}l \right).
  \label{eq:tangentTangentCorrelationsHB}
\end{equation}
A comparison of
\eref{eq:tangentTangentCorrelationsHB} with
\eref{eq:tangentCorrelationsGeneral} and identifying $l$ with $s$ results
in
\begin{equation}
  l_p=2\frac{Jb}{k_BT}=2\frac{\kappa}{k_BT}.
  \label{eq:persistenceLength}
\end{equation}
The persistence length is thus the ratio between bending stiffness
$\kappa$ and thermal energy $k_BT$ and is therefore a measure of the stiffness
of a polymer. In general dimension $d$ it holds \cite{LandauLifshitz}:
\begin{equation}
  l_p=\frac{2}{d-1}\frac{\kappa}{k_BT}.
  \label{PersistenceLength}
\end{equation}
There are three regimes defining three classes of polymers:
\begin{equation}
  \left \{
  \begin{array}{ll}
    b \approx l_p \ll L & \mathrm{flexible} \\
    b \ll l_p < L & \mathrm{semiflexible} \\
    b \ll L \ll l_p & \mathrm{stiff}. \\
  \end{array}
  \right.
  \label{eq:regimesOfTheWLC}
\end{equation}
At last we want to remark on the mean square end-to-end distance
$\langle R_{\rm ee}^2\rangle$.  Using the definition $\langle R_{\rm
ee}^2\rangle=\langle(b\sum_{i=1}^N\vec t_i)^2\rangle$ together with
\eref{eq:ratioOfBesselFunctions}, its calculation is straightforward and
amounts in the continuum limit to (cp.\ e.g.\ \cite{DoiEdwards}):
\begin{equation}
  \langle R_{\rm ee}^2\rangle=2l_pL\left
  \{1-\frac{l_p}{L}[1-\exp(-L/l_p)]\right \}.
  \label{eq:meanSquareEndToEndAnalytical}
\end{equation}

\subsection{Disorder}

The background potential consists of hard disks with diameter $\sigma$ that
interact with the monomers of the polymer via hard-core repulsion described by
the potential
\begin{equation}
  V = \left\{
  \begin{array}{l l}
    \infty & \quad \mathrm{for} \quad d < \sigma/2 \\ 0 &
    \quad \mathrm{else} 
  \end{array}
  \right. ,
\end{equation}
where $d$ is the distance between a monomer and a hard-disk center. Thus, the
monomers---here described by points---may not be placed onto the area of a
disk.

The assembly of the disks is the same as in \cite{Schoebl2011}. The disks are
put onto the sites of a square lattice with lattice constant $a$. Each site is occupied with a certain
occupation probability $p$ independent of the other sites. 
Consequently, there is no interaction between neighboring disks besides the constraint
that the minimum distance between two disk centers is $a$, the
lattice constant. This leads to clustering and hence a spatially
inhomogeneous structure of obstacles \cite{StaufferITPT}. 


%
\section{Algorithms}\label{sec:algorithm}

As in \cite{Schoebl2011}, we apply two algorithms for double-checking our
results. One is an off-lattice growth algorithm proposed by Garel and Orland
\cite{Orland1990} and one is the multicanonical Monte Carlo method [22--24].
Here, we only concentrate
on those aspects which are relevant for the semiflexible case.  Otherwise we
refer to \cite{Schoebl2011}. 

For the multicanonical approach we have developed in \cite{Schoebl2011} a
special modification, allowing us to reweight to different background potential
amplitudes.  To this end, we replace the infinite hard-disk potential with
finite potential steps and after performing the multicanonical simulation at
fixed persistence length, we are able to reweight to any potential amplitude
ranging from the free polymer (zero amplitude) to the polymer in a hard-disk
background (very large amplitude).

\begin{figure}%
  \hspace{-1cm}
  \begin{tabular}{c@{\hspace{-2cm}}c@{\hspace{-1cm}}c@{\hspace{-1cm}}cc}
  \includegraphics[scale=0.75]{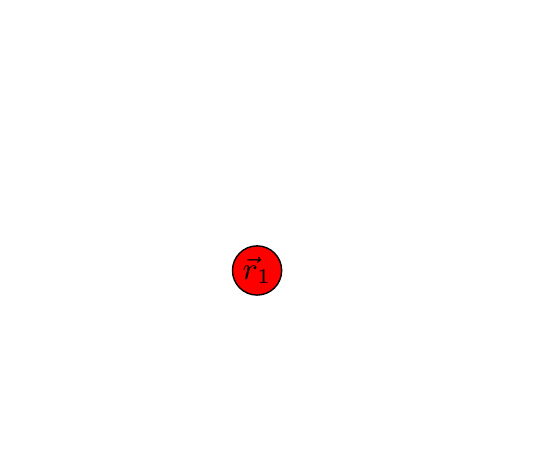}
  &
  \includegraphics[scale=0.75]{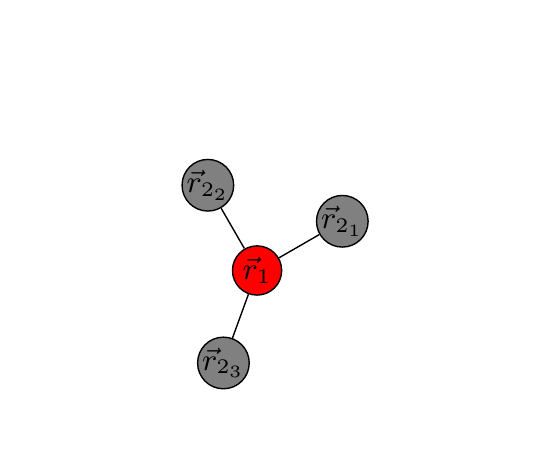}
  &
  \includegraphics[scale=0.75]{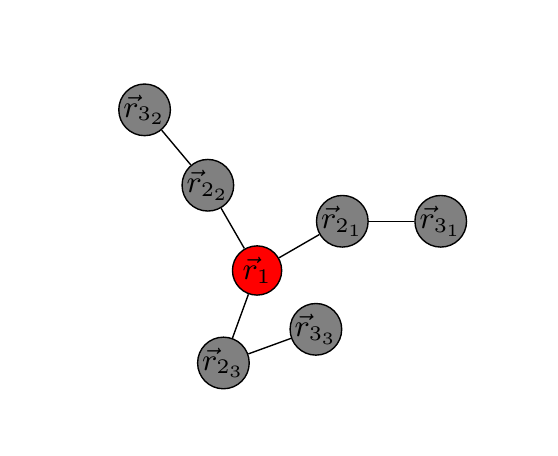}
  &
  \includegraphics[scale=0.75]{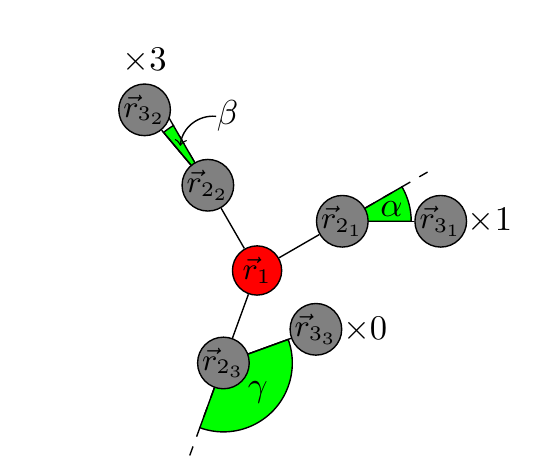}
  &
  \includegraphics[scale=0.75]{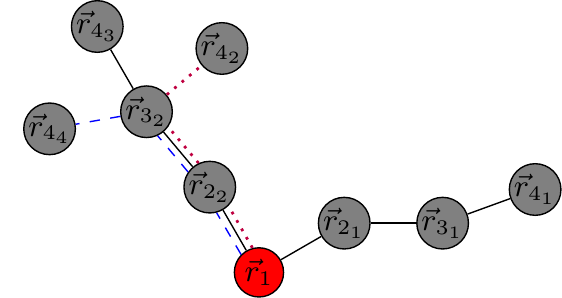}
  \\
  (a) & (b) & (c) & (d) & (e)
  \end{tabular}
  \caption{(Color online) (a) $\mathcal M_1$ monomers at position
  $\vec r_1$. The first monomer---here marked by \protect\tikz{\protect
  \draw[draw=red,fill=red] (0,0) circle (0.1cm);}---thus stands for $\mathcal
  M_1$ (three in this example) different chains of zero length.  
  \newline (b) Each of the $\mathcal M_1$ chains is extended by one monomer.
  There are now $M_2=3$ independent chains of length one. Up to now, there is no
  energy term as there is no bending angle between neighboring bonds.  
  \newline (c) Each of the $M_2$ chains is extended by one monomer. There are
  now $M_3=3$ independent chains of length two.
  \newline (d) Now, energy comes into play as there is a bending angle between
  the first and second bond of the polymers. Temperature $T$ and coupling constant
  $J$ are chosen such that they yield the weights that are given in the sketch
  ($\times 3,\times 1,\times 0$). Each of the chains is replicated according to
  its weight. Accordingly $M_{3_{\rm new}}=4$. There are now four independent chains
  of length two. 
  \newline (e) Each of these chains is extended independently by one monomer and bond. This
  procedure is iterated until the desired degree of polymerization is reached.
  }
\label{fig:basicPrincipleOfGrowthAlg}
\end{figure}

The basic routine of the growth method---also called replication-deletion procedure (RDP)---is
comprised in \fref{fig:basicPrincipleOfGrowthAlg}.  Polymers are grown in parallel from an
initial starting point. In each step, a polymer configuration is cloned according to the
Boltzmann weight $w_i$ for adding a new monomer. $i_i=\mathrm{Int}(w_i)$ is defined as the
integer part of $w_i$ and $r_i=w_i-i_i$ as the rest. Replicating the new chain $w_i$ times
statistically means replicating it $i_i$ times plus one additional time with probability
$r_i$. Therefore a random number $r$ with $0 \leq r \leq 1$ is drawn. If $r \geq r_i$, the
chain is replicated $i_i$ times. Otherwise it is replicated $(i_i+1)$ times.  Since $w_i$ can
be smaller than 1, the replication can in fact amount to a deletion. This is why the method is
called replication-deletion-procedure. The different clones are treated as independent polymer
configurations and are grown until the desired degree of polymerization is reached. In
dependence on the Hamiltonian of the system, cloning the configurations according to the
Boltzmann weight leads---except for very simple situations---to either exponential increase of
the numbers of configurations or to dying out of almost all configurations. Both cases defy the
estimation of meaningful numerical averages. This problem can be overcome by introducing a
population control parameter (PCP), ensuring that the number of sampled
configurations $\mathcal{M}_N$ roughly coincides with the initial number of
chains $\mathcal{M}_1$. For the principle of the PCP, we refer to the original
paper by Garel and Orland \cite{Orland1990} and to \cite{Schoebl2011}.

The RDP generates a population of chains that is Boltzmann distributed. To be
more precise, this procedure provides such a distribution in every single growth
step.  A strong advantage thereof is to be able to do a scaling analysis within
one simulation. Having a distribution of chains of length $N$ automatically
provides all the distributions of length $\tilde{N}=1,\ldots,N$. We found by
comparison with the multicanonical method that correlations due to the growth
process which would pass a possible bias from ensembles of short chains to those
of longer chains can largely be excluded. More on the correlations of chains
will be discussed in \sref{sec:averagingAndErrorEstimation}.

Although getting distributions of all lengths up to the desired degree of
polymerization within one simulation is an advantage for scaling analyses, it
might be a drawback concerning the question of ergodicity. Depending on the
choice of the potential, the polymer chain might, for example, get stuck in a
local energy minimum which hinders the chain from sampling phase space evenly
enough to provide a Boltzmann distributed population of chains that satisfies
the ergodicity condition. 

This drawback can be cured by introducing a guiding field that locally makes the
distribution of chains non-Boltzmann distributed thus facilitating to sample
phase space more uniformly by forcing the chain to circumvent or get out of
local energy minima \cite{Orland1990}. A second aspect of the guiding field is to make the
algorithm much more efficient.  Here, we bias the distribution of chains by
drawing angles not uniformly but from another distribution which is inspired by
the nature of the problem.  Afterwards, the weights have to be adapted such that the
resulting distribution is unbiased. The guiding field is made up of two parts,
one accounting for the bending energy of the polymer and one for the disks
of the background potential. 
\begin{figure}
  \centering
  \begin{tabular}{c@{\hspace{2cm}}c}
  \includegraphics[scale=0.75]{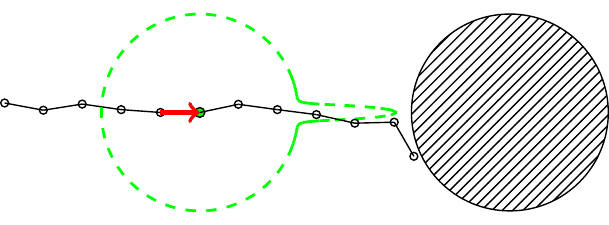}
&
  \includegraphics[scale=0.75]{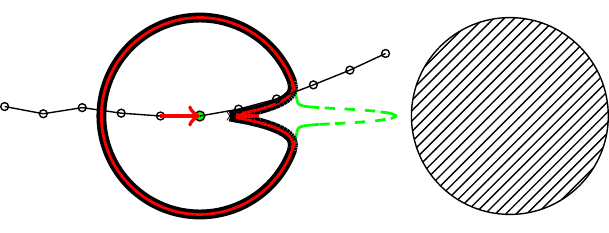}
\\
(a) & (b)
\end{tabular}
\caption{(Color online) Background-aware guiding field.}
\label{fig:guidingFieldBendingDisks}
\end{figure}

Assume a situation as sketched in \fref{fig:guidingFieldBendingDisks}(a), where
a polymer with a certain bending stiffness grows in the direction that is
indicated by the arrow (red). The hatched disk is an obstacle located in the
growth direction.  The dashed (green) line indicates the guiding field based
solely upon the bending stiffness. Both the guiding field and the Boltzmann
weight favor a growth in the direction of the bond indicated by the arrow and
thereby in the direction of the obstacle.  The polymer does not sense the
obstacle until it is one bond length away of it. It is obvious that only a large
bending angle can prevent the polymer from overlapping with the obstacle.
Depending on the bending stiffness, the resulting weight will be rather small
and the configuration does not contribute a lot or might even die out.  This
problem is based upon the update routine that only takes into account its
directly surrounding area. 

A way to overcome this problem is to introduce a guiding field that takes into
account the obstacles in the vicinity of the growing end of the polymer. Such a
guiding field is depicted in \fref{fig:guidingFieldBendingDisks}(b). The
probability of choosing an angle that leads in the direction of an obstacle is
reduced [framed (black) curve]. The corresponding probability density considers
only disks within a certain distance and adds for each disk a Gaussian dip with
a certain amplitude and variance. The form of the probability density and the
parameters are determined empirically and by intuition.  Both amplitude and
variance are a function of the distance between obstacle and monomer as well as
of the persistence length. The emerging growth direction is a
superposition of the contributions from the persistence of the polymer and from
the surrounding potential. It is evident that a polymer with a larger
persistence length has to sense the obstacles more in advance than one with a
smaller persistence length, because the probabilistic suppression of certain
angles depends exponentially on the bending stiffness. 

%
\subsection{Averaging and error estimation}\label{sec:averagingAndErrorEstimation}

We consider the background to be static on the timescale of polymer
fluctuations.  This is taken into account by performing the quenched disorder
average for calculating observables. Therefore two averages have to be carried
out. The first is an average over polymer configurations belonging to a single
disorder realization. It is written in angular brackets $\langle \ldots
\rangle$. This is done for all disorder realizations and the quenched average is
calculated thereof by averaging over the measured values of the single disorder
realizations. The polymer configurations that belong to a single disorder
realization are all pinned at the same pinpoint.  Leaving the constraint of the
pinpoint is discussed in \sref{sec:leavingConstFixedEnd}. The quenched average
is written as $[\langle ...  \rangle]$.  

Consequently, two kinds of variances have to be considered, one from the average
of polymer configurations within a single disorder realization, and the other
from the average over different disorder realizations.  These two contributions
amount in an effective variance $\sigma_{\rm eff}^2$ which is estimated by (see
e.g.\ \cite{ErrorEstimation}):
\begin{equation} \sigma_{\rm eff}^2 = \frac{\sigma_{\mathcal
  O}^2}{N_{r}},
  \label{eq:averageIndependentMeasurements} 
\end{equation}
where $\sigma_{{\mathcal O}}^2$ is the variance of the Monte Carlo mean values
over a finite sample of $N_{r}$ different (independent)
disorder realizations. For the error bar we take the standard deviation
$\sqrt{\sigma_{\rm eff}^2}$. For the case of fixed pinpoints, the quenched
average is carried out over $N_{r}=1500$ independent
disorder realizations. With this statistical precision, the relative error turned out to be
of the order of $1\%$, which is far smaller than the effect of the disorder on
the observables. For the scales considered here, the error bars are covered by
the plot markers.  Therefore we omit them. 

In order to achieve a reasonable balance between the amount of computing time
invested in the polymer statistics for a given disorder realization and the
number of independent disorder realizations, at least a rough estimate of the
statistical error of the polymer simulations is needed.  The estimation of this
error for a simulation within a single disorder realization for the case of the
multicanonical Monte Carlo method is well described in literature, e.g.\
\cite{ErrorEstimation}. Things are more complicated for the case of the growth
algorithm. There, the different polymer configurations cannot be assumed
independent. If we recall the principle of the algorithm, we realize that many
polymers share a certain part of their configuration which leads to correlations
in the final ensemble. Once having found the number of independent
configurations, the error can be estimated after
\eref{eq:averageIndependentMeasurements}.  For the free polymer, we follow the
approach of Higgs and Orland in \cite{HiggsOrland1991}. They estimated the
variance by assuming that interactions are only between nearest neighbors. As
the free polymer model (no disorder) within this work only includes bending
energy between neighboring bonds, it fulfills the preconditions of the error
estimation by Higgs and Orland. Under this assumption they found the number of
independent chains $c_{\rm ind}$ to be proportional to $\mathcal M_1/N$, where
$\mathcal M_1$ is the initial number of chains and $N$ the number of bonds.  The
variance of the simulation of a free single chain is calculated by applying
\eref{eq:averageIndependentMeasurements} with $\sigma_{\rm eff}^2$ substituted
by the variance of the mean value of a single simulation $\tilde
\sigma_{\overline {\mathcal O}}^2$ and $\sigma_{\mathcal O}^2$ substituted by
the fluctuations of the chains belonging to a single simulation $\tilde
\sigma_{\mathcal O_j}^2$. The number of realizations $N_{r}$ is substituted by the
independent number of chains which---according to
\cite{HiggsOrland1991}---yields:
\begin{equation} 
  \tilde \sigma_{\overline {\mathcal O}}^2\sim\frac{\tilde \sigma_{\mathcal
  O_j}^2}{\mathcal M_1/N},
  \label{eq:averageIndependentMeasurementsSingle} 
\end{equation}
where $\tilde \sigma$ indicates the case of a single simulation without disorder
and without quenched average. The error bars are again taken to be the
standard deviation calculated from
\eref{eq:averageIndependentMeasurementsSingle}.  If we add disorder, the
estimation of the number of uncorrelated configurations becomes more difficult
as the narrow channels between neighboring disks, especially for high area
occupation probabilities, bring about additional correlations. We assessed the necessary
number of polymer chains for producing averages in appropriate accuracy by
considering the mean values for increasing number of chain configurations. 

For all ranges of disorder occupation and persistence length, we found the
maximum relative deviations of the mean values for $\mathcal M_1=50000$ and
$\mathcal M_1=100000$ to be about $5\%$ while the relative deviations for
$\mathcal M_1=100000$ and $\mathcal M_1=400000$ are only about $1\%$. As the
deviations between the latter two are much below the effect of the influence of
the disorder, the accuracy obtained by simulating with $\mathcal
M_1=100000$ is completely satisfactory for the scope of this work. The above
estimation is reassured by a crosscheck with a completely different method---the
multicanonical Monte Carlo method. 

\section{Observables, parameters, and test cases}\label{sec:obsParamsTestCases}

\subsection{Observables}\label{sec:observables}

Throughout this work we focus on two observables: the end-to-end distribution
$P(r)$ and the tangent-tangent correlations 
$\langle \vec{t}(0)\vec{t}(kb) \rangle$
The end-to-end distribution gives the probability to find
a certain \mbox{end-to-end} distance $r=b|\sum_{i=1}^N \vec t_i|$. 
The tangent-tangent correlation function $\langle \vec{t}(0)\vec{t}(kb) \rangle$ is
estimated by averaging the mean tangent-tangent correlation functions of a single 
polymer configuration over all sampled configurations $N_{p}$ ($\mathcal M_N$
for the growth algorithm)
\begin{equation}
  \langle \vec{t}(0)\vec{t}(kb) \rangle
   = \frac{1}{N_{p}} \sum_{j=1}^{N_{p}}\left( \overline{\vec{t}_0\vec{t}_{k}} \right)_{j}
   = \frac{1}{N_{p}} \sum_{j=1}^{N_{p}}\left( \frac{1}{N-k}\sum_{i=1}^{N-k}\vec{t}_i\vec{t}_{i+k} \right)_{j}.
\end{equation}
The tangent-tangent correlation
function is a measure of the stiffness of a polymer.  For a completely flexible
free polymer there is no energetic preference to any angle and hence there are
no correlations between tangent vectors for $k\neq0$.  For the case with bending
stiffness, the tangent correlations are described by
\eref{eq:ratioOfBesselFunctions}. The surrounding disorder can lead to both
correlations and anti-correlations [as can be seen later in
\fref{fig:singleDisorderConfigurationAnalysis}(b)].

\subsection{Simulation parameters and length scales}\label{sec:simulationParameters}

The polymer determines three length scales of the system. The total length $L$,
the persistence length $l_p$, and the bond length $b$.  The former two are
reduced to the ratio $\xi=l_p/L$, which is the persistence length measured in
units of polymer length $L$. The persistence lengths considered here include
$\xi=0$, representing the flexible case, and $0.1,0.2,0.3,0.5,0.7,1$. The
contour length $L$ and the bond length $b$ are related by $Nb=L$, so that $b$
resp.\ $N$ sets the scale of discretization. In our case, the discrete polymer
model has $N=29$ bonds which corresponds to $N+1=30$ monomers. As stated in
\sref{sec:polymerModel}, the polymer is a phantom chain, i.e., there is no
steric self-interaction of the chain (the monomers are considered pointlike).
The issue of discretization is touched in \sref{sec:freePolymer}.

The simulations are done in a square box with periodic boundary conditions
filled with a $20\times20$ lattice with lattice constant $a$. We consider the
site occupation probabilities $p=0, 0.13, 0.25, 0.38, 0.51, 0.64,
0.76, 0.89, 1.00$. The occupation probabilities are specified to be
consistent with those from \cite{Schoebl2011}, where they were chosen to equal
the area fractions $\rho=0,0.1,\cdots,0.7,0.785$ for $\sigma=a$. The diameter
of the disks here is set to $\sigma=0.9a$, which introduces a small channel
between neighboring disks. The issue of $\sigma=a$ and $\sigma>a$ is briefly
discussed in \sref{sec:impactDiskDiameter}.  Unless otherwise stated, the
numerical results refer to the case of $\sigma=0.9a$. 

The disorder brings another two length scales into play. One is the disk
diameter $\sigma$, another is the average free distance between the centers of the
disks $l_0$. $l_0$ is connected to the occupation probability $p$ via
$l_0=a/\sqrt{p}$, where $a$ is the lattice constant. For our parameter choice
($\sigma=0.9a$) this amounts to 
\begin{equation}
  l_0=\frac{1.11\sigma}{\sqrt{p}}.
  \label{eq:averageFreeDistance}
\end{equation}  
Note that $l_0$ does not account for the extension of the disks.
The top of \tref{tab:averageFreeDistance} gives an overview over $l_0$ in dependence on the
occupation probability $p$. 

The length scales of the polymer and those of the disorder are connected via
$L=6.4\sigma$ or, equivalently, $\sigma=4.5b$ (for our choice $N=29$). Note that $a=5b$, which
amounts to an effective distance of half the bond length between neighboring disks.
For a better comparison of the length scales, the bottom of
\tref{tab:averageFreeDistance} shows the persistence length $l_p$ of a free
polymer in units of $\sigma$.

The simulation parameters are chosen such that we can investigate both the
effect of the smallest structures and the impact of the disorder on the polymer
on the length scale of several disk diameters $\sigma$.
Going to much larger chains at the same accuracy involves a much higher
computational effort. The effects we are looking at would, however, be
qualitatively the same.

\begin{table}
  \caption{\label{tab:averageFreeDistance}Top: Average distance between the
  centers of the disks in dependence on the occupation probability $p$. Bottom:
  Persistence length and root mean square end-to-end distance of a free polymer
  in units of $\sigma$ in dependence on $\xi$.}
  \begin{indented}
  \item[]
    \begin{tabular}{ccccccccc} 
      \br
      $p$ & 0.13 & 0.25 & 0.38 & 0.51 & 0.64 & 0.76 & 0.89 & 1.00 \\ 
      \mr
      $l_0$ & $3.1\sigma$ & $2.2\sigma$ & $1.8\sigma$ & $1.5\sigma$ &
      $1.4\sigma$ & $1.3\sigma$ & $1.2\sigma$ & $1.1\sigma$\\
      \br
    \end{tabular}
  \item[]
    \begin{tabular}{ccccccc} 
      \br
      $\xi$ & 0.1 & 0.2 & 0.3 & 0.5 & 0.7 & 1.0\\ 
      \mr
      $l_p$ & $0.64\sigma$ & $1.3\sigma$ & $1.9\sigma$ & $3.2\sigma$ &
      $4.5\sigma$ & $6.4\sigma$\\
      $\sqrt{\langle R_{ee}^2\rangle}$ & $2.7\sigma$ & $3.6\sigma$ & $4.2\sigma$ &
      $4.9\sigma$ & $5.2\sigma$ & $5.5\sigma$\\
      \br
    \end{tabular}
  \end{indented}
\end{table}

\subsection{Test cases} 

\subsubsection{The free polymer}\label{sec:freePolymer}

The free semiflexible polymer is already widely discussed throughout literature
[17, 27--31].
Here we will just mention
some characteristics as the free case will always serve as reference for the
case with disorder. 
\begin{figure}
  \hspace{-1.3cm}
  \begin{tabular}{c@{\hspace{-0.0cm}}c}
    \includegraphics{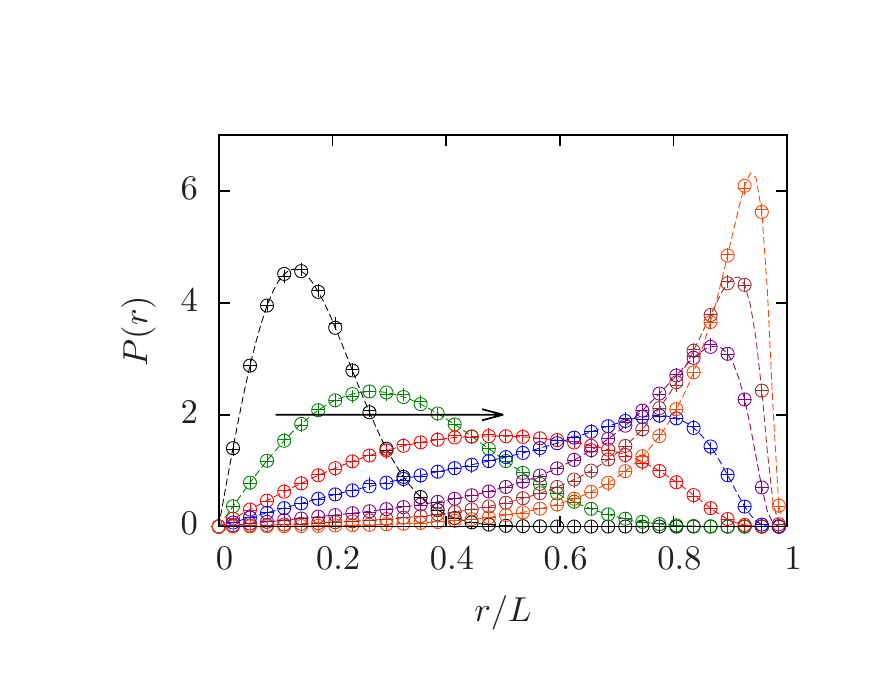}
    &
    \includegraphics{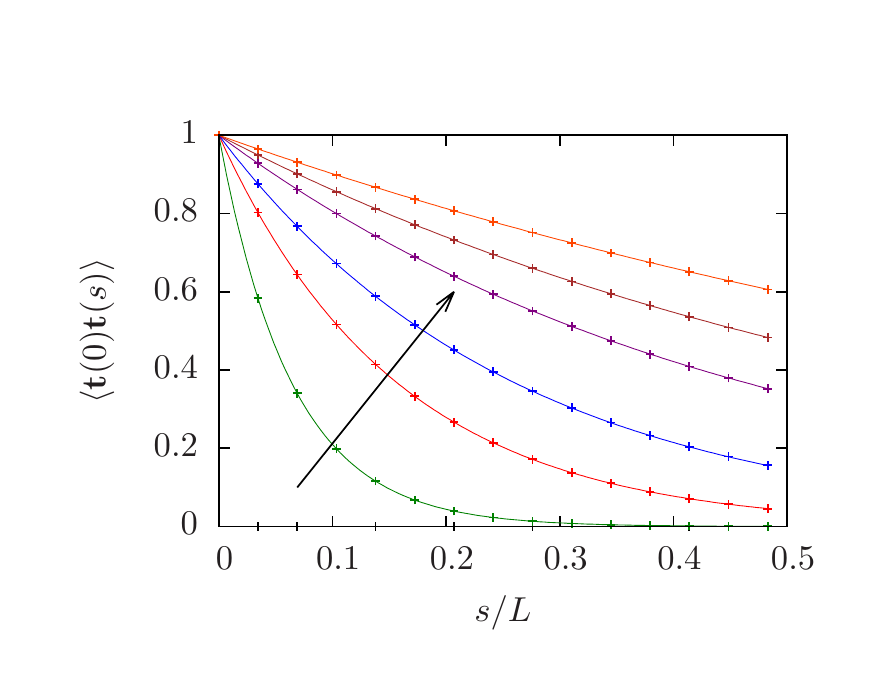}
    \\
    (a) & (b) 
  \end{tabular}
  \caption[]{\label{fig:freePol}(Color online) End-to-end distribution function (a) and tangent
  correlation function (b) of free semiflexible polymers with $N+1=30$ monomers. The persistence
  lengths include $\xi=0,0.1,0.2,0.3,0.5,0.7,1$ (increasing persistence
  indicated by the arrow).\\(a): $\circ$ are data from the growth method; $+$
  are Metropolis data. The connecting lines are drawn for better
  visibility.\\(b): The solid
  lines are the analytical solution of the tangent-tangent correlations
  \eref{eq:ratioOfBesselFunctions}. The trivial case of $\xi=0$---immediate
  decorrelation---is not shown.}
\end{figure}
\Fref{fig:freePol} shows the measured observables from \sref{sec:observables}---the
radial distribution function $P(r)$, \fref{fig:freePol}(a), and the
tangent-tangent correlations $\langle \vec{t}(0)\vec{t}(s) \rangle$,
\fref{fig:freePol}(b). The functional form of the end-to-end distribution
function $P(r)$ of the free polymer is characterized by a single peak whose
position depends on the stiffness of the polymer. The probability of extended
chain configurations increases with increasing stiffness. Hence the peak is
shifted to the right for increasing bending energy. The tangent-tangent
correlations are shown in \fref{fig:freePol}(b) and cover the solid lines from
\eref{eq:ratioOfBesselFunctions} perfectly.  For the case of no persistence, the
tangent-tangent correlation function drops immediately to zero as there is no
correlation between the bonds besides the trivial self-correlation at $s=0$.

\begin{figure}[t]
  \centering \includegraphics[scale=1]{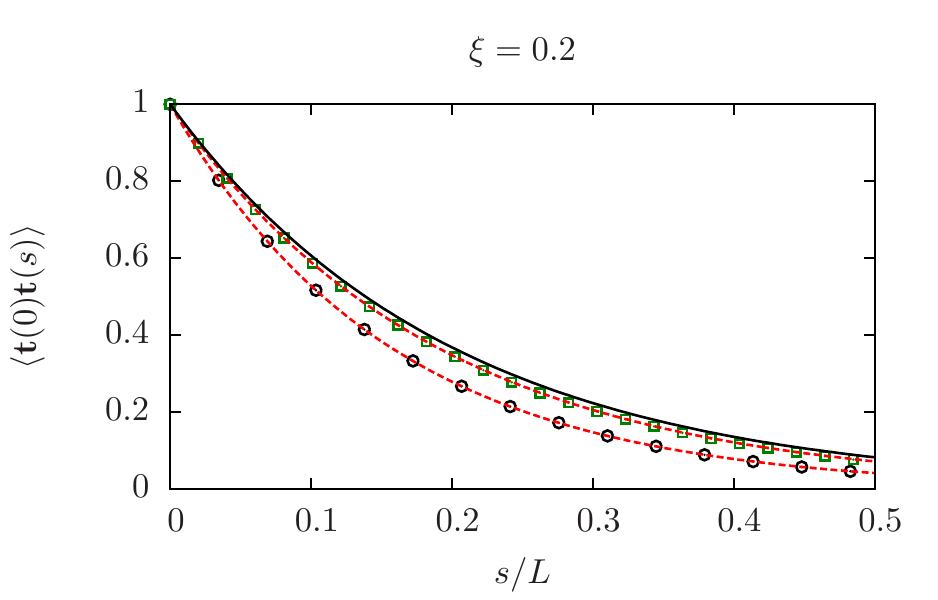} \caption[Discretization
  effects shown for the tangent-tangent
  correlations]{\label{fig:freePolTanTanCorrP02growthAlgAnalyt}(Color online)
  Tangent-tangent correlation function of a free semiflexible polymer.  {\tiny
  $\bigodot$} for the case of 30 monomers and {\tiny $\boxdot$} for the case of
  100 monomers.  The dashed lines show the analytical solution
  \eref{eq:ratioOfBesselFunctions} of the tangent-tangent correlations for the
  discrete case. The solid line shows the analytical solution
  \eref{eq:tangentCorrelationsGeneral} resp.\
  \eref{eq:tangentTangentCorrelationsHB} for the continuous case.} 
\end{figure}

An important aspect for comparison with analytical work on the worm-like
chain model is the degree of discretization of the polymer.
\Fref{fig:freePolTanTanCorrP02growthAlgAnalyt} shows the tangent-tangent
correlation function of a free semiflexible polymer at $\xi=0.2$. The deviations
from the continuous case are shown.  In the limit of small $b$ or large $\beta
J$, and therefore large $N$, the continuous case---exponential decay of the
tangent-tangent correlations \eref{eq:tangentTangentCorrelationsHB}---is
recovered.  

\subsubsection{Single disorder configuration}

\begin{figure}[h]
  \begin{tabular}{c@{\hspace{-0cm}}c} 
    \multicolumn{2}{c}{\includegraphics[scale=0.3]{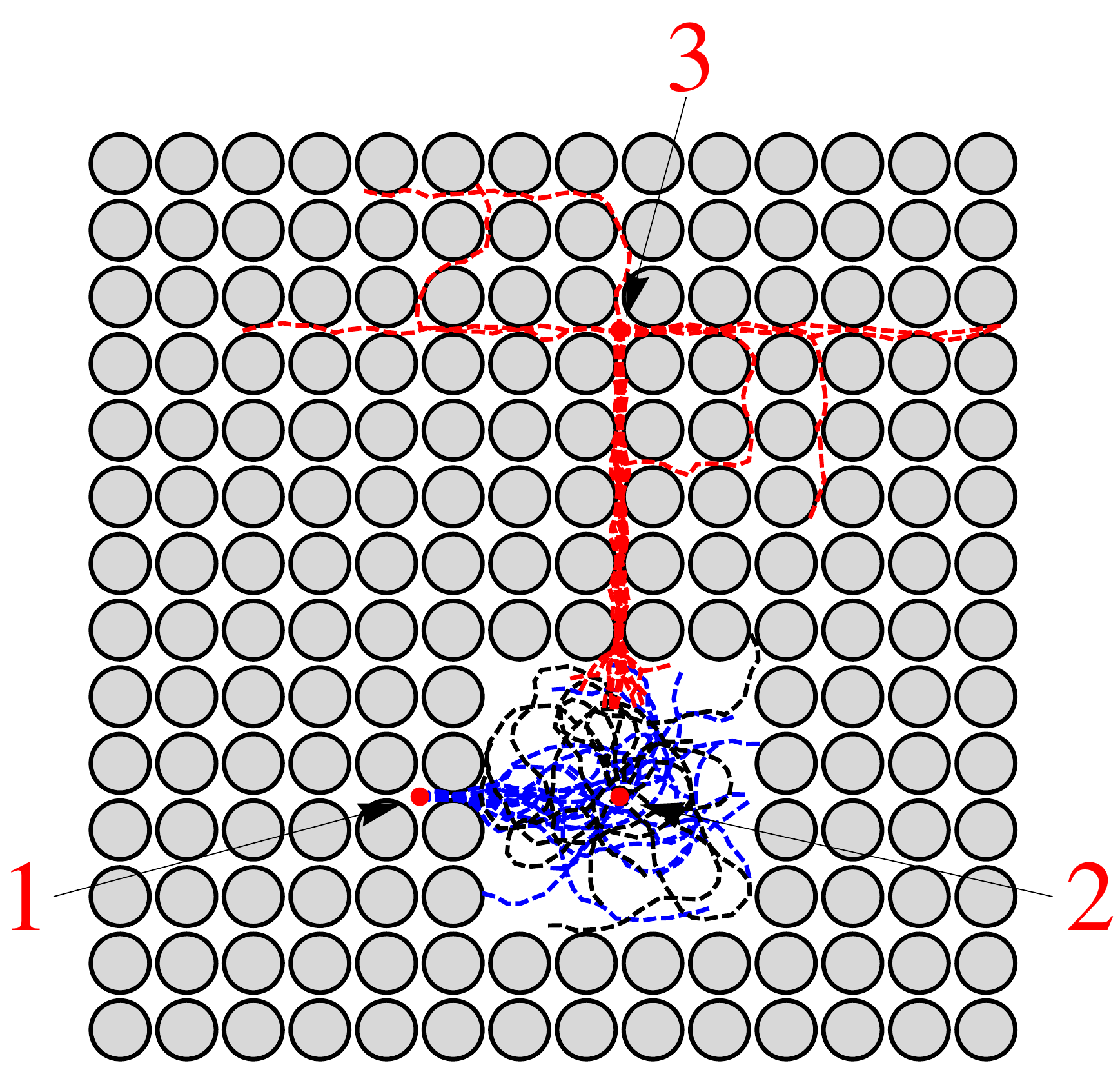}}
    \\
    \hspace{-1.5cm}
  \includegraphics[scale=1]{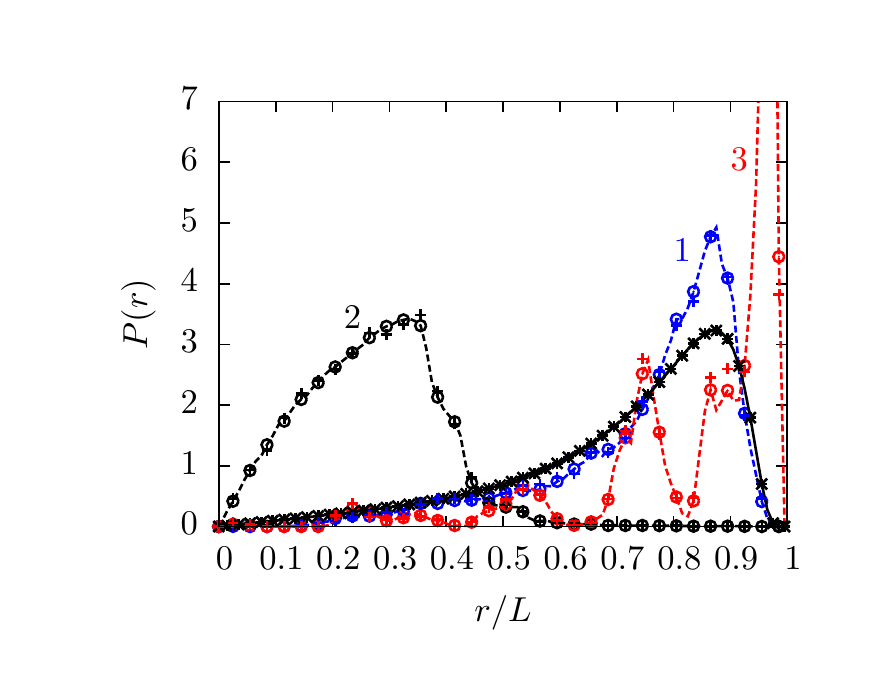}
  &
    \hspace{-1.5cm}
  \includegraphics[scale=1]{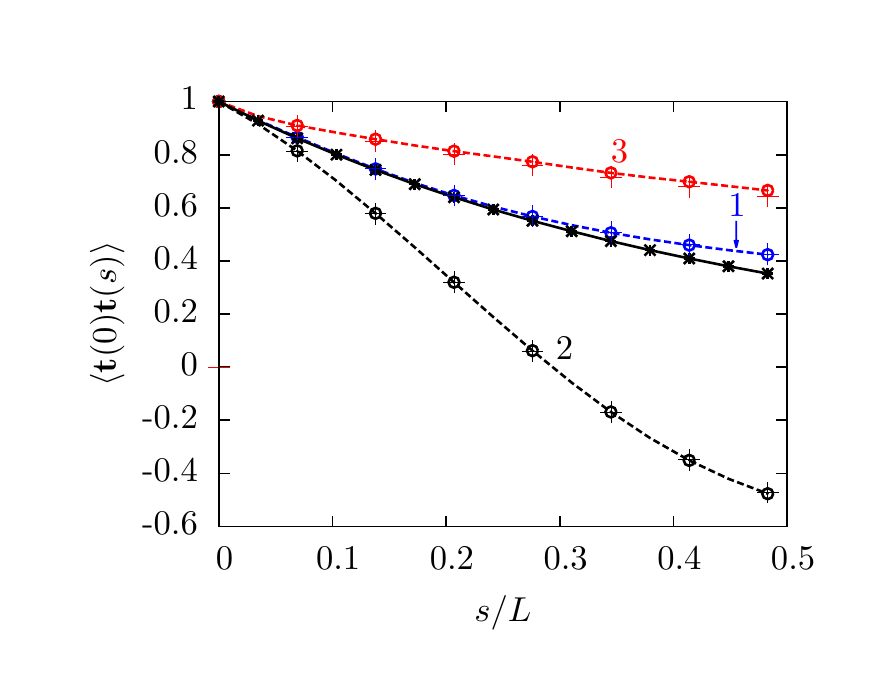}
    \\
    (a) & (b) 
  \end{tabular}
    \caption{\label{fig:singleDisorderConfigurationAnalysis}(Color online) Top: Distribution of
    disks with three exemplary pinpoints. The sketch additionally shows a
    selection of strongly contributing polymer configurations. \\Bottom: End-to-end
    distribution (a) and tangent-tangent correlations (b) that belong to the
    pinpoints shown above (single simulation; no disorder average) for $\xi=0.5$.  $\circ$
    shows the data from the growth algorithm; $+$ are data from the
    multicanonical simulation. The labeling of the curves is
    given in the plot. The curve marked by $\ast$
    shows the free case. The connecting lines are drawn for
    better visibility.}
\end{figure}

We are now adding obstacles to the system. Before we look at the quenched
average, we consider three exemplary pinpoints within an artificial disorder
configuration, where all sites are occupied except a $4\times4$ square.
\Fref{fig:singleDisorderConfigurationAnalysis} illustrates
the case with persistence paradigmatically for $\xi=0.5$.  
We start by discussing the issue of pinpoint~1.  While the only determining
factor for the case without bending energy was entropy, energy gains more and
more importance as soon as we start to increase the persistence length.  The
gain in entropy for configurations pinned to pinpoint~1 by exploring the large
free area around pinpoint~2 favors configurations that reach to that space.
Going straight through the channel from pinpoint~1 to the free area is even
forwarded by the energetic preference for small bending angles.
\Fref{fig:singleDisorderConfigurationAnalysis}(a) and (b) demonstrate that both
the end-to-end distribution function and the tangent-tangent correlation
function for the case of pinpoint~1 are similar to the free polymer.  This is
reasonable as the space available for the polymer to spread, once having passed
the narrow channel from pinpoint~1 to the adjacent region, provides entropically
similar space as for a free polymer. Space for bending back is strongly limited
by the potential but as this is energetically not opportune anyway, it does
barely affect the equilibrium ensemble. This behavior changes if we move on to
configurations starting from pinpoint~2.  While this pinpoint provided good
preconditions for a flexible polymer to behave as its free counterpart (see
\cite{Schoebl2011}), a free polymer with $\xi=0.5$ has a mean extension of about
$4.9\sigma$ (cp.\ \tref{tab:averageFreeDistance}).  The free space in
each direction from pinpoint~2 is about $2\sigma$ which truncates a large part
of configuration space. While the confinement forces the polymer to crumple up,
the energetic cost for bending stretches the polymer out. The interplay of these
effects leads to the formation of loops with strong anticorrelations on the
length scale of the persistence length, which is half the polymer length [see
\fref{fig:singleDisorderConfigurationAnalysis}(b)]. Although configurations
starting from pinpoint~1 are similar to the free case, configurations starting
from pinpoint~2 are ``flexibilized''.  In contrast, configurations starting from
pinpoint~3 show the very reverse---stiffening by disorder. The energetic drive
to stretch the polymer allows for finding favorable spots even if they are far
away which is just opposite to the flexible case. The large area around
pinpoint~2 facilitates a spread of the chain which leads to a strong entropy
gain.  Therefore, the equilibrium ensemble is strongly dominated by
configurations that end in the large free space.  Accordingly, the end-to-end
distribution is peaked around almost completely stretched configurations and the
bonds are strongly correlated on all lengths, which can be seen in
figures~\ref{fig:singleDisorderConfigurationAnalysis}(a) and (b). Some less
distinct peaks stem from configurations that are kinked once, twice, etc. The
contribution of those configurations quickly decreases with increasing number of
kinks, because each kink strongly increases the bending energy. 

Now that we have investigated different scenarios that can occur during the
quenched average and thus gained insight into some dominating elements of the
quenched average, we move on to averaging over many disorder realizations.

\section{Results}\label{sec:quenchedAverage}

In \cite{Schoebl2011} the crossover between a low- and a high-density regime is
determined by the occupation $p_0$ where the mean end-to-end distance of the
polymer equals the average distance between neighboring disks. This estimation
works better for the case without persistence because the polymers for the case
with bending stiffness are more extended in linear shapes, especially for large
persistence lengths. We will see that the high-density regime is characterized
by a multiple peak structure in the \mbox{end-to-end} distribution function. The
beginning of this effect shows up as a small bulge in the end-to-end
distribution function. This effect even increases for the case with persistence.
\begin{figure}[b]
  \centering \includegraphics[]{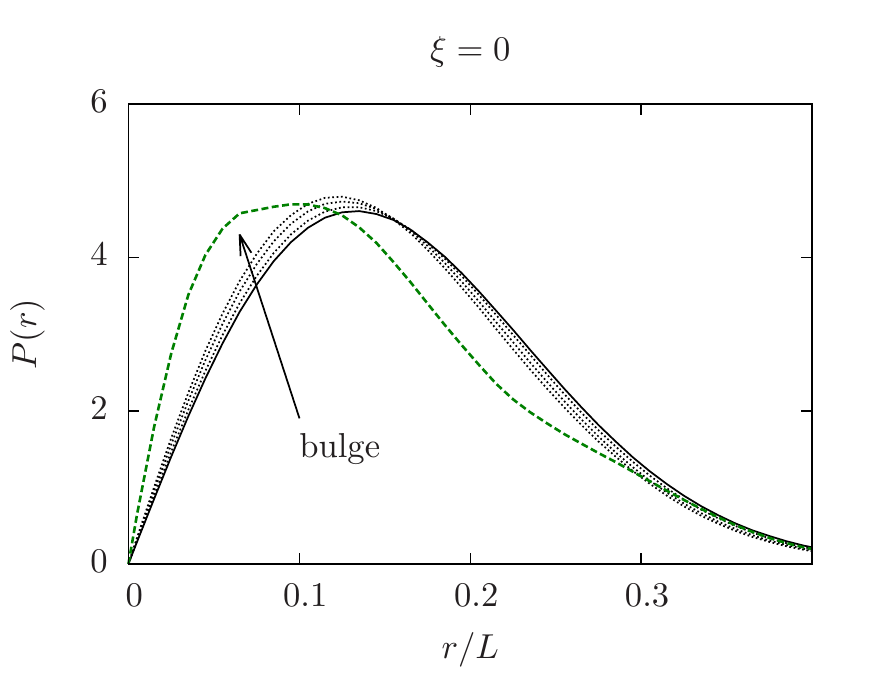}
  \caption{\label{fig:indicatorCrossover}(Color online) End-to-end distribution
  function for the case of no persistence. The occupation probabilities are
  $p=0,0.13,0.25,0.38$, and $0.76$ (indicated by the arrow). The solid and
  dotted (black) curves are in the low-density regime. The high-density
  case $p=0.76$, labeled by the long-dashed (green) curve, is indicated by a deviation
  of the functional form from the $p=0$ (black solid) case.}
\end{figure}
\Fref{fig:indicatorCrossover} shows the end-to-end distribution function of a
flexible polymer in hard-disk disorder. The solid and dotted (black) curves are
in the low density regime. The low-density distributions all have the same
functional form which is characterized by a single peak. The dashed (green)
curve for high-density disorder differs from this structure---it has a bulge.
The shape of the bulge that emerges as soon as a certain occupation $p_0$ is
crossed can be used as indicator to mark the crossover from a low-density to a
high-density regime.  Similar effects can be observed for the tangent-tangent
correlations. We take the qualitative change of the functional form of the
end-to-end distribution---formation of double/multiple peak structure---as a
qualitative signal for the crossover from a low- to a high-density regime.  

\subsection{Low-density regime}
\begin{figure}
  \hspace{-1.5cm}
  \begin{tabular}{c@{\hspace{-0.0cm}}c}
    \includegraphics{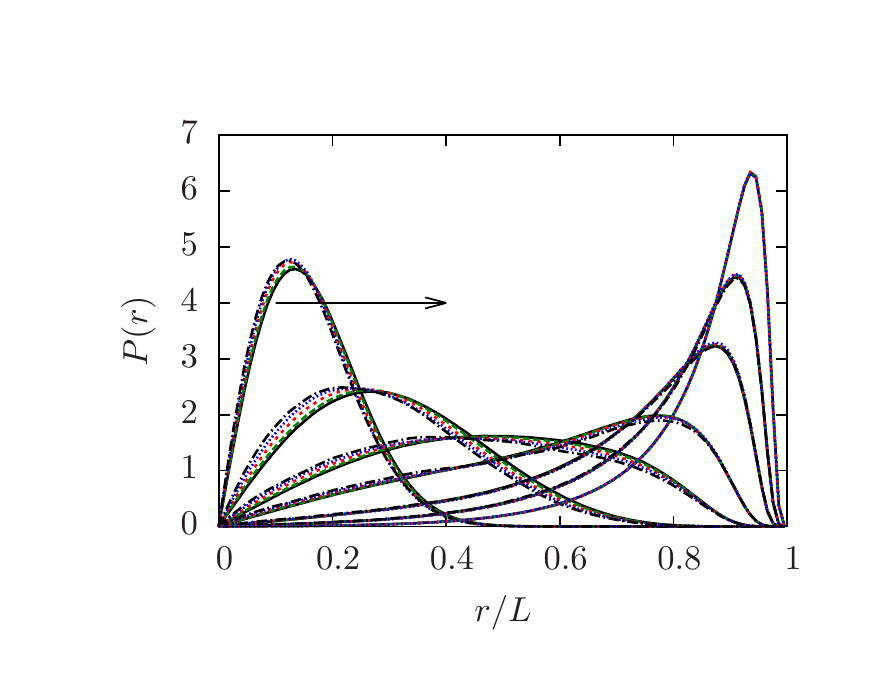}
    &
    \includegraphics{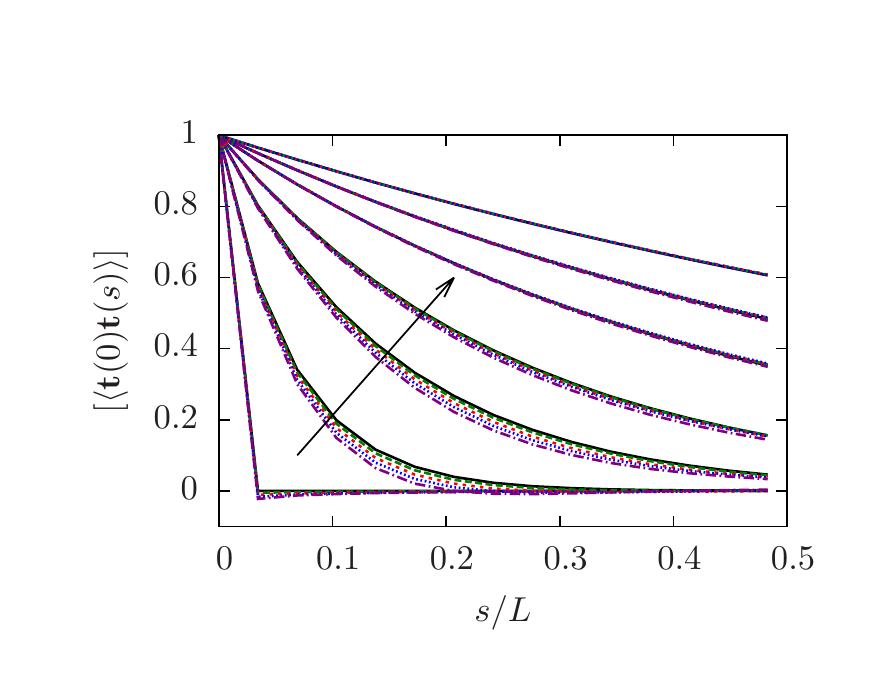}
    \\
    (a) & (b)
  \end{tabular}
  \caption{\label{fig:lowDensityRegime}(Color online) End-to-end distribution (a) and
  tangent-tangent correlations (b) in
  the low-density regime. The persistence lengths
  include \mbox{$\xi=0,0.1,0.2,0.3,0.5,0.7,1$} (indicated by the
  arrow).  The occupation probabilities are $p=0$
  (\hspace{-0.2cm}\protect\includegraphics[scale=1]{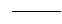},
  black), $0.13$
  (\hspace{-0.2cm}\protect\includegraphics[scale=1]{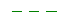},
  green), $0.25$
  (\hspace{-0.2cm}\protect\includegraphics[scale=1]{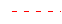},
  red), $0.38$
  (\hspace{-0.2cm}\protect\includegraphics[scale=1]{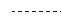},
  blue), and $0.51$
  (\hspace{-0.2cm}\protect\includegraphics[scale=1]{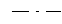},
  black).}
\end{figure}

In the low-density regime, the relevant parameter is the persistence length and not the 
disorder density.
\Fref{fig:lowDensityRegime} shows the measured data for increasing persistence
(growing $\xi$ is indicated by the arrow). The occupation probabilities in the
plot are $p=0,0.13,0.25,0.38,0.51$, corresponding to an average distance between
the disks of $l_0\geq 1.5\sigma$ (cp.\ \tref{tab:averageFreeDistance}). 
We find two kinds of response to the disorder depending on the stiffness of the
polymer---compression and extension.  The probability for shorter end-to-end
distances is growing for increasing occupation $p$ at low persistence lengths
$\xi\le0.2$ corresponding to $l_p\le1.3\sigma$ which is less than the smallest
average mean distance $l_0=1.5\sigma$ within this density regime. The reverse is
observed for $\xi\ge0.5$ which corresponds to $l_p\ge3.2\sigma$ which is more
than the largest average distance $l_0=3.1\sigma$ (except for $p=0$) in the
low-density regime (the effect of stiffening is hardly seen in
\fref{fig:lowDensityRegime}). Remember that higher probability for shorter
chains in the end-to-end distribution function [\fref{fig:lowDensityRegime}(a)]
and faster decay of the tangent correlations [\fref{fig:lowDensityRegime}(b)]
indicate softening. The reverse effect is analogous.  

\begin{figure}
  \centering
  \begin{tabular}{c@{\hspace{1.5cm}}c@{\hspace{1.5cm}}c@{\hspace{1.5cm}}c}
    $\xi$ small & $\xi$ large & $\xi$ small & $\xi$ large \\
    \includegraphics[scale=0.24]{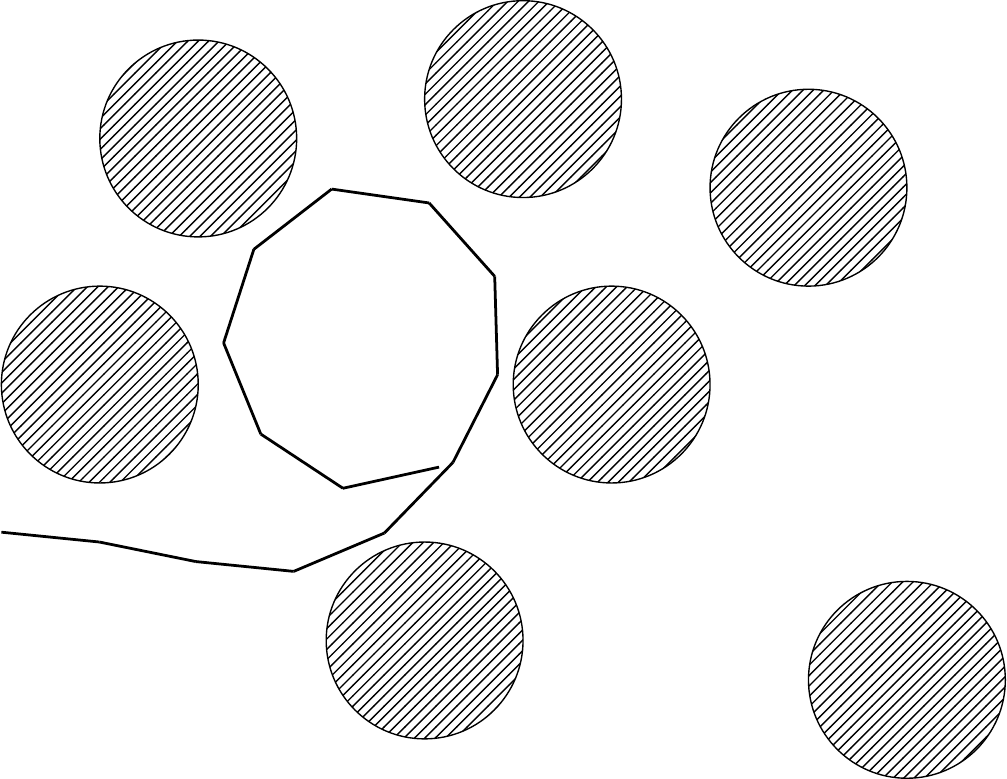}
    &
    \includegraphics[scale=0.24]{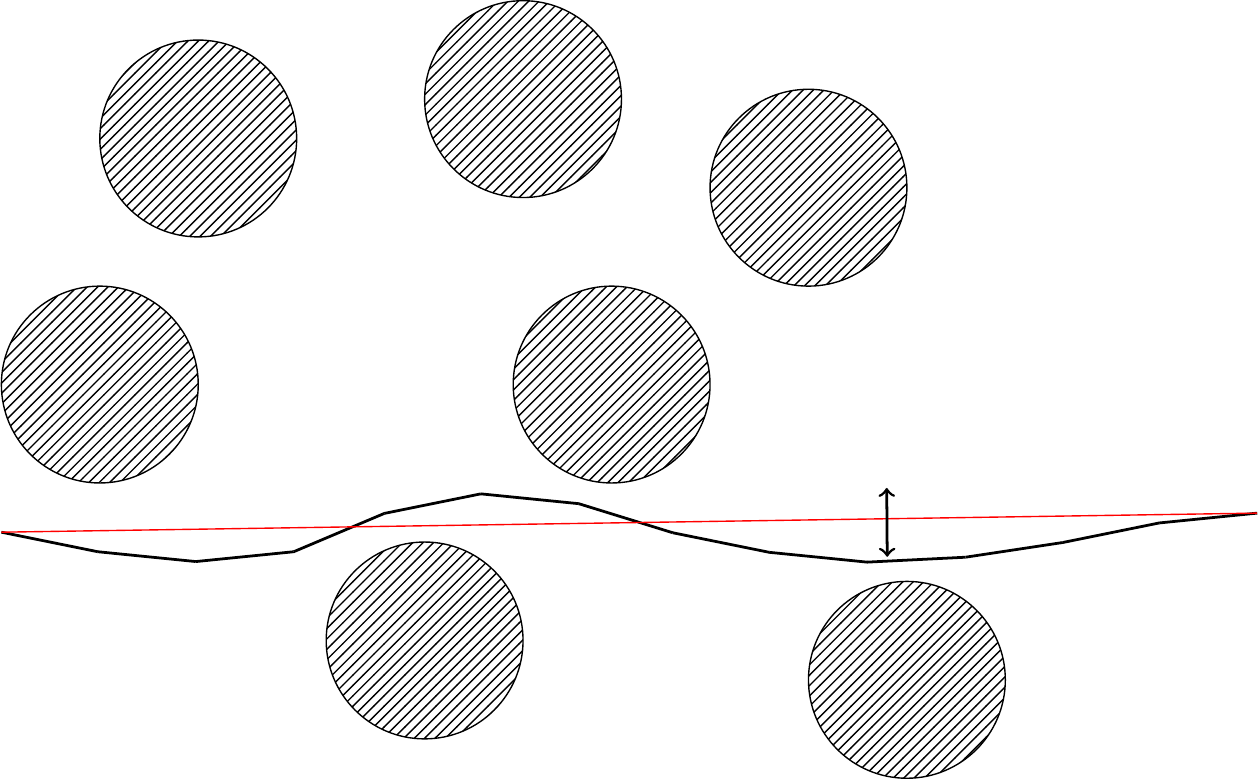}
    &
    \includegraphics[scale=0.375]{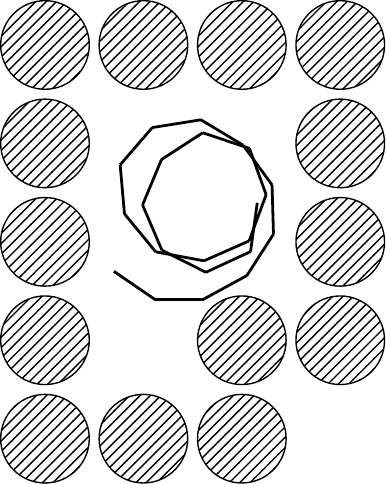}
    &
    \includegraphics[scale=0.375]{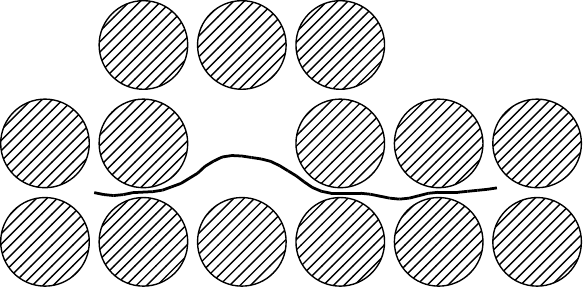}
    \\
    \multicolumn{2}{c}{\hspace{-2cm} (a)} & \multicolumn{2}{c}{\hspace{-1.5cm} (b)}\\
  \end{tabular}
  \caption{\label{fig:sketchPolymerDisorder}(Color online) Sketch to elucidate
  the idea of softening and stiffening for persistent polymers at low (a) and
  high (b) occupation probabilities, respectively. The double-headed arrow
  indicates the width of the thermal fluctuations of the polymer.}
\end{figure}

For an explanation of the softening and stiffening at different persistence
lengths consider \fref{fig:sketchPolymerDisorder}(a). The case of small
persistence lengths is shown on the left in \fref{fig:sketchPolymerDisorder}(a).
The energetic cost for bending is in a range where it is more favorable for the
polymer to crumple up in order to gain entropy than to stretch. This is
different for stiffer polymers. The probability for bending decreases exponentially
with increasing persistence.  That is why configurations are favored that find
tube-like free regions. The width of thermal fluctuations of those
configurations is limited by the distance between neighboring disks
[right in \fref{fig:sketchPolymerDisorder}(a)]. The squared width of the
fluctuations is related
to the persistence length via (refer to, e.g., \cite{1367-2630-9-11-416})
\begin{equation}
  \frac{\delta r_{\perp}^2}{L^2} \propto \frac{1}{\xi},
  \label{eq:fluctWidth}
\end{equation}
i.e., it decreases for increasing persistence length. A large persistence length
hence corresponds to a small fluctuation width. Thus, in the limit of large
bending stiffness, the stiffening effect induced by disorder vanishes. 

\subsection{High-density regime}

We now turn over to the high-density regime with $p\geq0.64$ (which is above the
percolation threshold $p_c=0.5927$) where the shape of the distributions
starts to exhibit characteristics of the potential up to the point where the
potential completely dominates the distributions. This means that the
confinement increases in such a way that the polymer either has to crumple up
even though this is connected to high cost in energy or has to stretch at the
expense of entropy.  

We consider the effect of high-density disorder for three
exemplary persistence lengths, $\xi=0.1$, $0.3$, and $1$.  $\xi=0.1$
represents a quite flexible polymer that can well adapt to the surrounding
disorder by crumpling up. $\xi=1$ is rather stiff with respect to the disorder
and adapting to confinement by crumpling is only feasible at high energetic
cost. In this case, adapting is mostly done by stretching. $\xi=0.3$ is in
between and exhibits both, crumpling and stretching. The influence of the disk
diameter is discussed at the end of this section. 
\begin{figure}[t]
  \hspace{-1.3cm}
  \begin{tabular}{c@{\hspace{-2.5cm}}c@{\hspace{-2.5cm}}c}
  \includegraphics{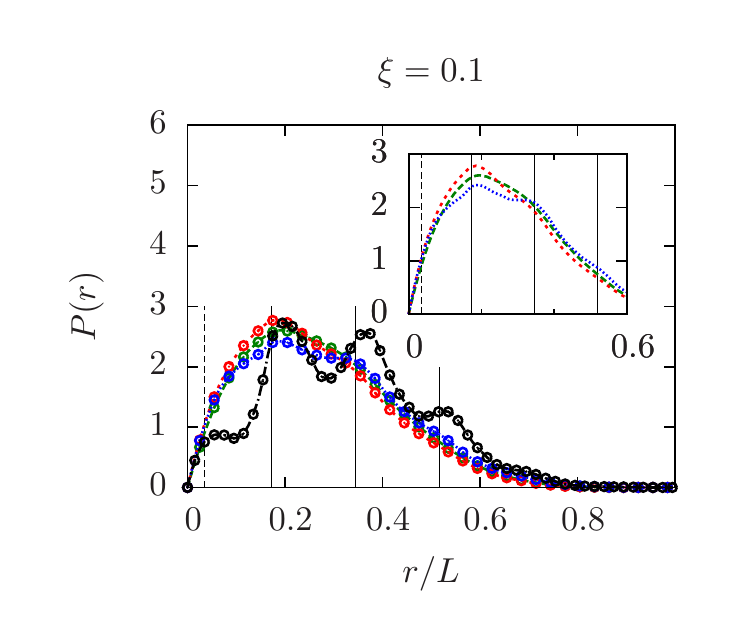}
  &
  \includegraphics{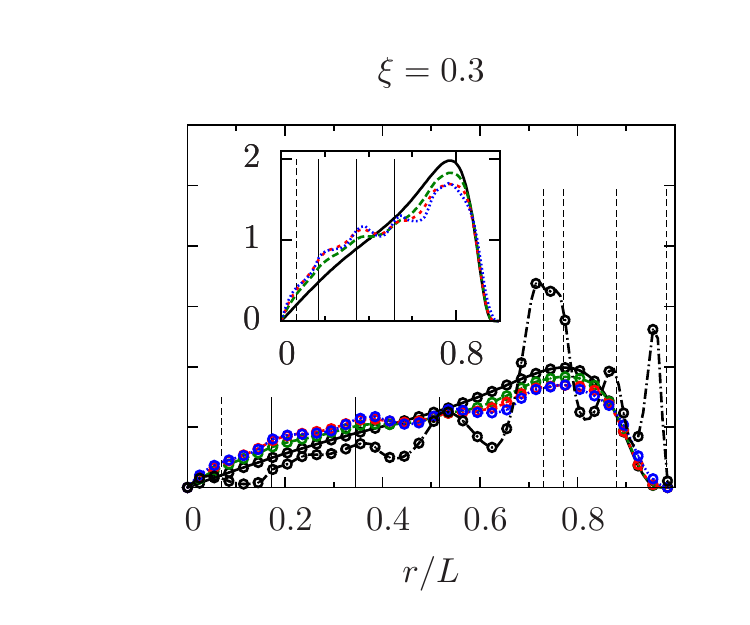}
    &
  \includegraphics{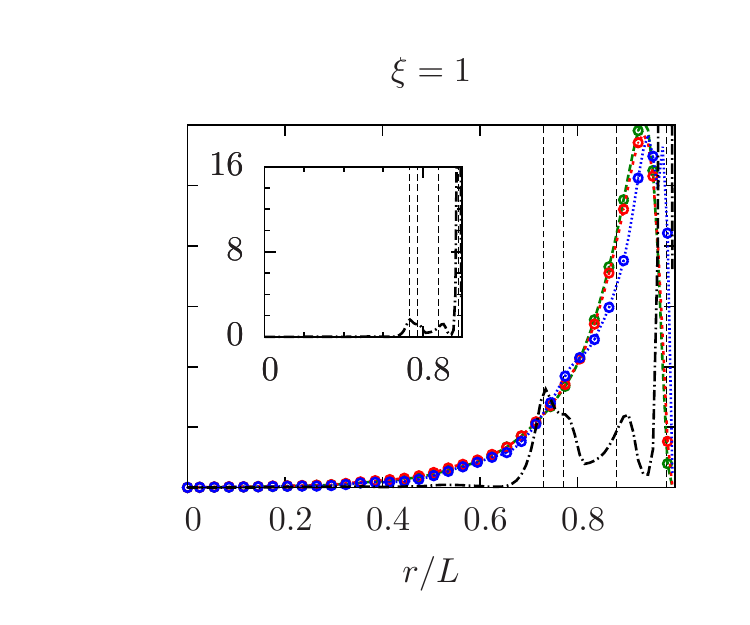}
  \vspace{-1.3cm}
  \\
  \includegraphics{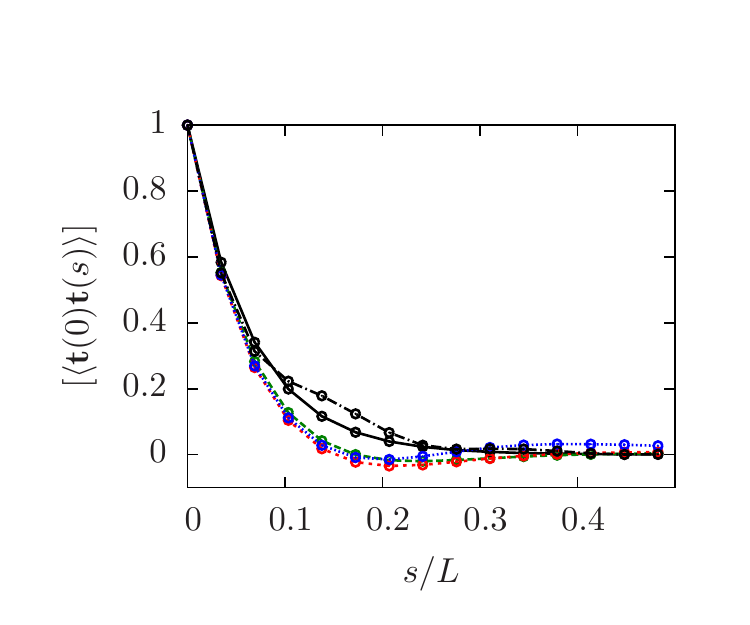}
  &
  \includegraphics{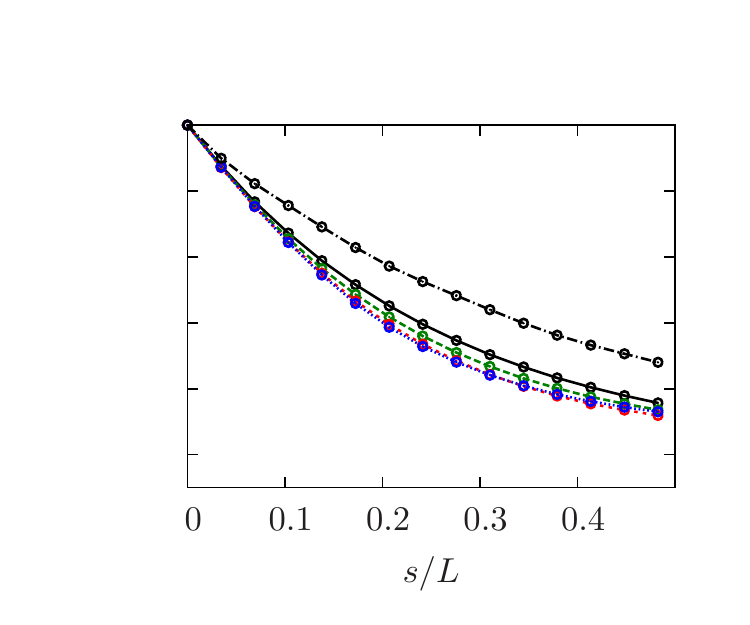}
    &
  \includegraphics{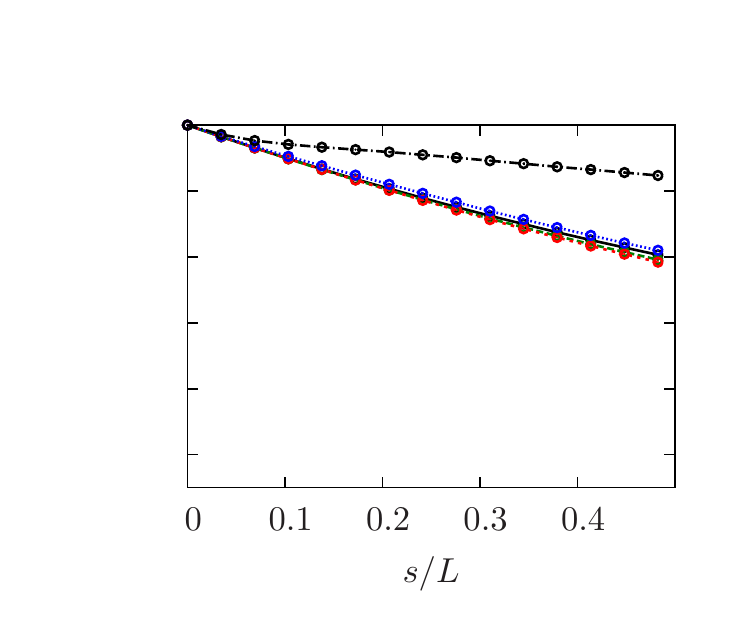}
  \\
  \hspace{1cm} (a) & \hspace{1cm} (b) & \hspace{1cm} (c)
  \end{tabular}
  \caption{\label{fig:endToEndHigh}(Color online) End-to-end distribution function (top) and
  tangent correlation function (bottom) for $\xi=0.1$ (a), 0.3 (b), and 1 (c).
  The occupation probabilities are: $p=0$
  (\hspace{-0.2cm}\protect\includegraphics[scale=1]{blackSolid.pdf},
  black), $p=0.64$
  (\hspace{-0.2cm}\protect\includegraphics[scale=1]{greenLongDashed.pdf},
  green), $0.76$
  (\hspace{-0.2cm}\protect\includegraphics[scale=1]{redDashed.pdf}, red), $0.89$
  (\hspace{-0.2cm}\protect\includegraphics[scale=1]{blueDotted.pdf}, blue), and
  $1$ (\hspace{-0.2cm}\protect\includegraphics[scale=1]{blackDashDotted.pdf},
  black). The vertical lines in the end-to-end distribution functions correspond
  to the distances shown in \fref{fig:gridDistances}.}
\end{figure}
In contrast to the low-density regime, the distributions in the high-density
regime feature a variety of peaks due to the confining effect
of the potential. The periodic structure of the lattice is mirrored in the
observables that characterize the polymers. 

\subsubsection{Small persistence length}

The end-to-end distribution and the tangent-tangent correlations for $\xi=0.1$
are shown in \fref{fig:endToEndHigh}(a). As long as the persistence length is of
the order of the extension of the available space, the polymer crumples up close
to its pinpoint. 
\begin{figure}
  \centering
  \begin{tabular}{m{0.4\linewidth}m{0.4\linewidth}}
    \includegraphics[scale=0.4]{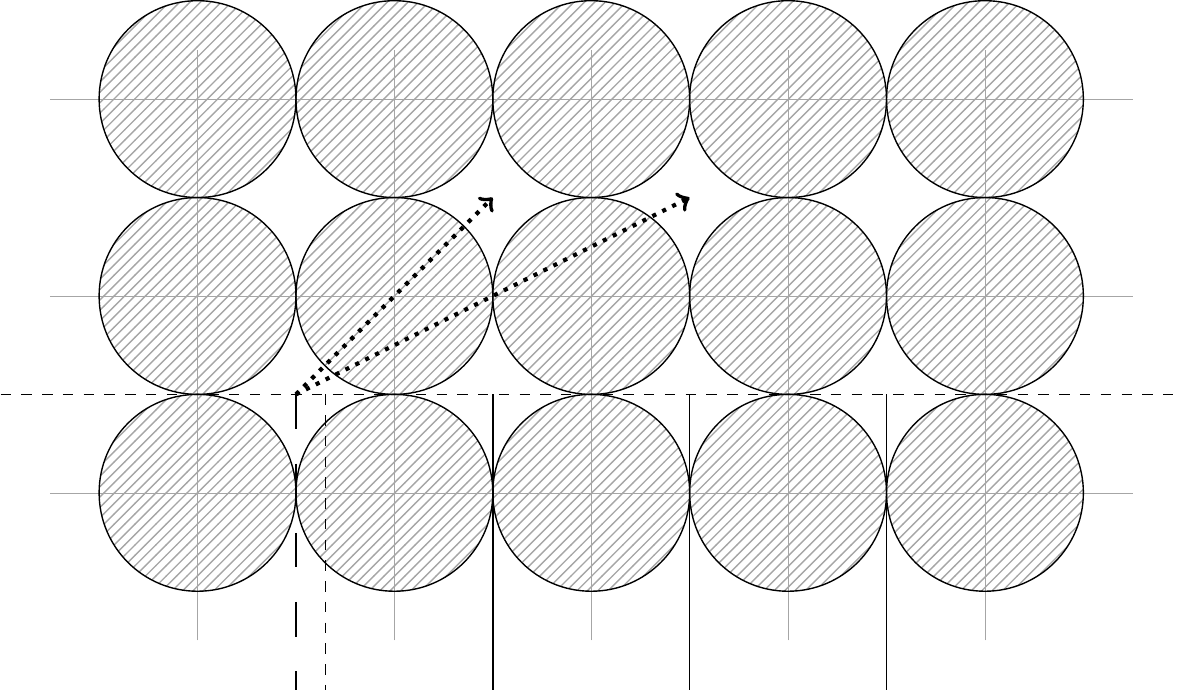}
    &
    \includegraphics[scale=0.4]{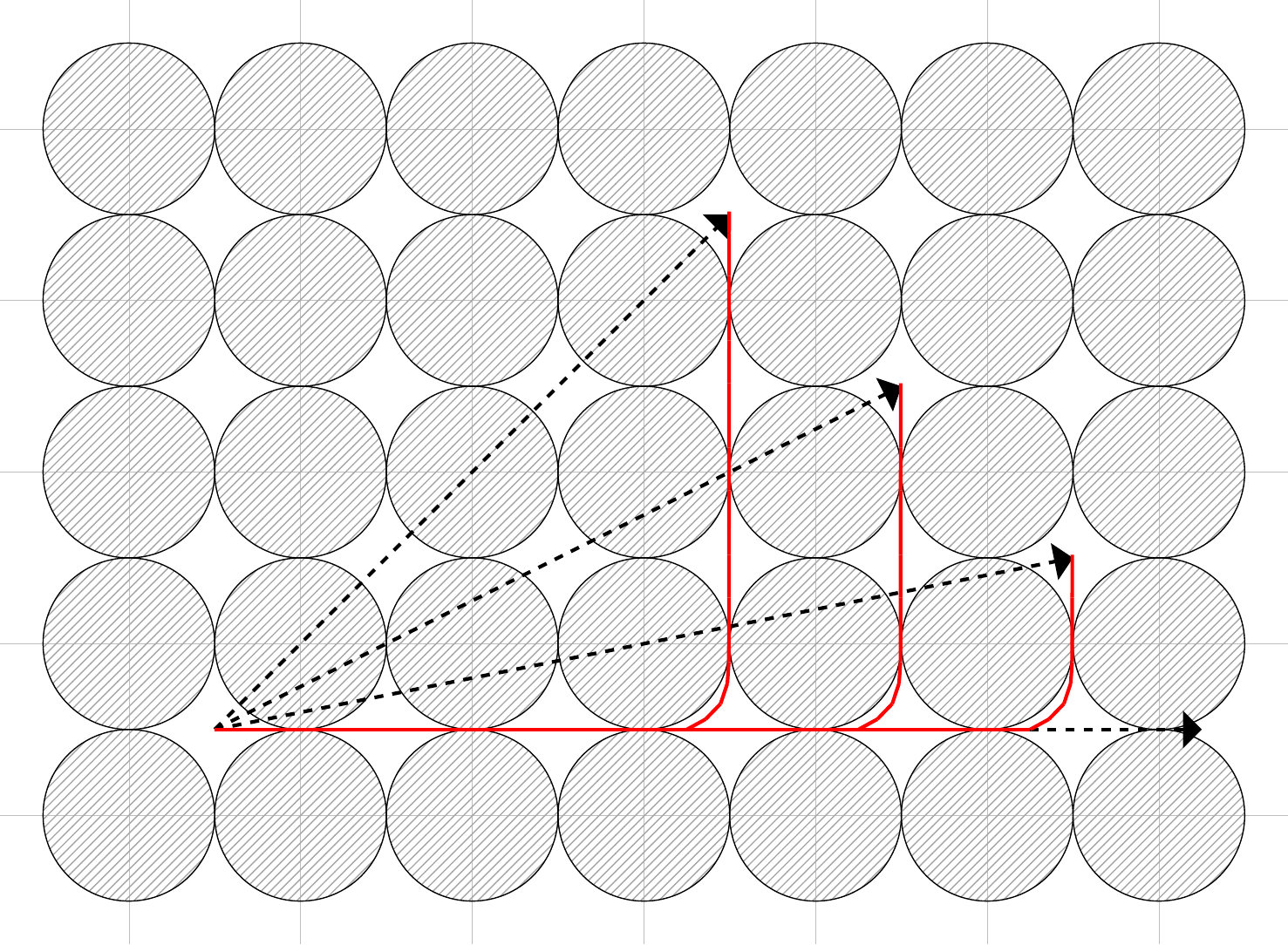}
    \\
    \hspace{2cm}(a) & \hspace{2.5cm}(b)
  \end{tabular}
  \caption{\label{fig:gridDistances}Section of a lattice. The part shown here is
  fully occupied, which is just exemplary. (a): The long-dashed vertical line
  (left most) is a reference line. The other lines and arrows show different
  distances on the lattice. The short-dashed vertical line stands for the mean
  extension in a small cavity. The next vertical line depicts the distance one
  lattice constant $a$ apart, the next $2a$, and so on. (b): The horizontal
  arrow indicates the end-to-end distance of fully stretched polymers which are
  prevailing for large $p$ and $\xi$. The other arrows are end-to-end distances
  to cavities that are reached by polymers with one $90^{\circ}$ turn. Three of
  them are indicated by the dashed (red) lines.}
\end{figure}
The persistence length $\xi=0.1$ corresponds to three bonds. This is of the
order of the extension of the smallest cavities (like those around pinpoint~$1$ in
\fref{fig:singleDisorderConfigurationAnalysis}; we call them $\diamond$-cavities
as they are shaped like a diamond $\diamond$).
The left part of \fref{fig:sketchPolymerDisorder}(b) illustrates this situation. Crumpled
configurations are reflected in the contributions to small extensions in the
end-to-end distribution function [top of \fref{fig:endToEndHigh}(a)]. Another
indicator is a sharper decline of the tangent-tangent correlations [bottom of
\fref{fig:endToEndHigh}(a)]. The peaks in the end-to-end distribution become
more pronounced with increasing occupation probability $p$.  Some configurations
(the fraction of those increases with increasing $p$) will extend to
neighboring regions once the energetic penalty for bending becomes too large or it is
entropically more favorable for the polymer to extend to neighboring cavities,
respectively. The latter situation plays a crucial role especially for flexible polymers \cite{Schoebl2011,YamakovMilchev1997,echeverria:104902}.

As soon as the $\diamond$-cavities contribute the dominant part to the starting
points, especially for the case of $p=1$, the peaks in the end-to-end
distribution function can directly be ascribed to the periodic structure of the
lattice. \Fref{fig:gridDistances}(a) shows the different length scales that
mainly determine the extension of the polymer in this case. Large clusters of
void space do not play a role in this regime. The polymer, starting at one
point, will either stay near the region where it started or extend through a
channel to a neighboring or next-nearest neighboring, etc., free region. The
first distance, indicated by the short-dashed vertical line in
\fref{fig:gridDistances}(a), plays a role for very high occupation probabilities
($p=0.89,1$), as most of the chains will start in a small cavity.  The lines of
\fref{fig:gridDistances}(a) are sketched in \fref{fig:endToEndHigh} (lines left
of 0.6). The dotted lines (arrows) play a subordinate role and are therefore
omitted in figures~\ref{fig:endToEndHigh} and \ref{fig:impactDiskDiameter}. The reason is that a polymer that moves on to a neighboring
cavity instead of staying in the current one has lower energy if it goes straight,
which is not the case for the cavities indicated by the arrows. Their role
becomes even less important with increasing bending stiffness. 

\subsubsection{Large persistence length}

Next we are looking at the stiff counterpart. \Fref{fig:endToEndHigh}(c) shows
the case of $\xi=1$, which is a typical representative of semiflexible polymers [cp.\
\eref{eq:regimesOfTheWLC}], where bending on the length scale of a few bonds is
punished by high energetic cost.  The end-to-end distribution function for
$\xi=1$ also exhibits the periodic structure which is preset by the structure of
the potential. It is, however, much less pronounced and most of the
contributions stem from extended chains.  The right of
\fref{fig:sketchPolymerDisorder}(b) is an illustration of a polymer with a
persistence length that is larger than the average void-space cluster size. Some
configurations will still crumple up in small cavities, which, however, make
only a vanishing small contribution. Extended chains contribute the most part.
\Fref{fig:sketchPolymerDisorder}(b) (right) shows a rather extended
configuration. Some end-to-end length is stored in a cluster of size one in an
undulation. As soon as the lattice is fully occupied, the width of the
transverse fluctuations are strongly suppressed. Additionally, the polymer
behaves like a stiff rod on the length scale of the $\diamond$-cavities.
Accordingly, extended configurations prevail in this regime and the end-to-end
distribution function is dominated by a single peak near one. The only
additional significant contributions stem from configurations which are kinked
once. The end-to-end distances belonging to those configurations are sketched in
\fref{fig:gridDistances}(b). The distances belonging to these configurations are
indicated in \fref{fig:endToEndHigh}(c). 

\subsubsection{Crossover}

The end-to-end distribution function and the tangent-tangent correlations for
the intermediate stiffness with $\xi=0.3$ are shown in \fref{fig:endToEndHigh}(b).
The free polymer, indicated by the solid (black) line, has a peak at quite
extended configurations.  The persistence length counted in numbers of bonds is
about $9$, which is larger than the extension of the
$\diamond$-cavities.  Stretching is promoted by energy and by the channel
structure of the potential. On the other hand, the confinement, especially the
channel structure at $p=1$, reduces configuration space thus being unfavorable
with respect to entropy.  

The transition from $\xi=0.1$, which is rather flexible, to the quite stiff case
of $\xi=1$ via the intermediate stiffness of $\xi=0.3$ is well seen for $p=1$.
While $\xi=0.1$ has no contributions to extended chains and $\xi=1$ has none to
coiled configurations, $\xi=0.3$ has both [see \fref{fig:endToEndHigh}(top)]. The
distances of \fref{fig:gridDistances}(a) and (b) are sketched. The lines do not
match as nicely as in the case of $\xi=1$ because a smaller persistence length
allows larger amplitudes of undulations and hence smaller end-to-end distances.
The average effect is comprised in the tangent-tangent correlations [bottom of
\fref{fig:endToEndHigh}(b)] which reveals that the chain has all in all become
stiffer. 

The end-to-end distribution and tangent-tangent correlations for other
persistence lengths are not shown here as they are a composition of the effects
that contribute to the rather flexible case of $\xi=0.1$ and the much stiffer
case of $\xi=1$ as we have seen for $\xi=0.3$. 

\subsection{Impact of the disk diameter}\label{sec:impactDiskDiameter}

Similar to the approach in \cite{Schoebl2011}, we also investigated the impact
of the disk diameter $\sigma$ on the polymer distributions. An increase of
$\sigma$ to $\sigma=a$ leaves only pointlike channels between neighboring disks.
As the choice of our model only forbids overlaps of the monomers (but not of the
bonds) with the disks of the potential, the polymers for the case of $\sigma=a$
can still cross these channels. Crossing such a narrow channel leads, however,
to a strong decrease of entropy. Hence this is only favorable if a large void
space is reached by doing so or by balancing the entropy drawback by an energy
benefit in having fairly stretched configurations. Consequently, the effects
found above are enhanced and more pronounced. 
\begin{figure}
  \centering
  \begin{tabular}{c@{\hspace{-2.5cm}}c}
    \includegraphics{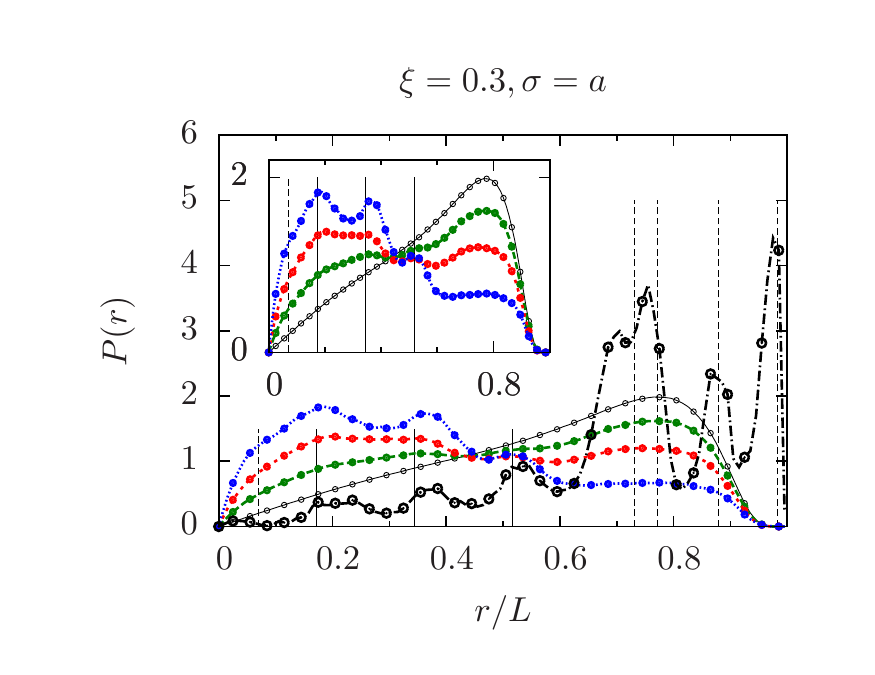}
    &
    \includegraphics{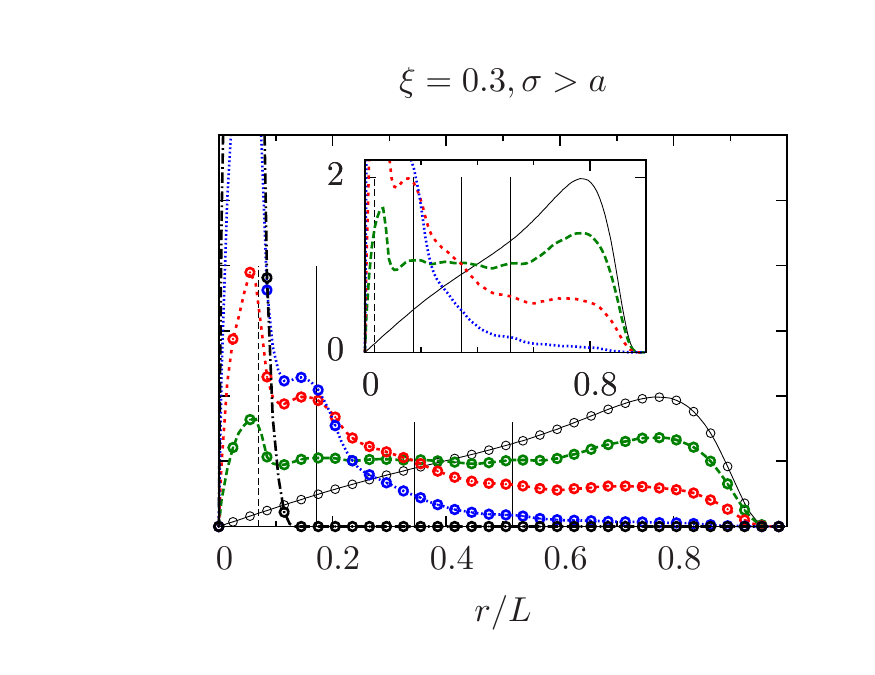}
    \\
    (a) & (b)
  \end{tabular}
  \caption{\label{fig:impactDiskDiameter}(Color online) End-to-end distribution for $\sigma=a$
  (a) and $\sigma>a$ (b). The occupation probabilities are $p=0$
  (\hspace{-0.2cm}\protect\includegraphics[scale=1]{blackSolid.pdf}, black),
  $0.64$ (\hspace{-0.2cm}\protect\includegraphics[scale=1]{greenLongDashed.pdf},
  green), $0.76$
  (\hspace{-0.2cm}\protect\includegraphics[scale=1]{redDashed.pdf}, red),\\* $0.89$
  (\hspace{-0.2cm}\protect\includegraphics[scale=1]{blueDotted.pdf}, blue), and
  $1$ (\hspace{-0.2cm}\protect\includegraphics[scale=1]{blackDashDotted.pdf},
  black).}
\end{figure}
The impact of the disk diameter is illustrated for the example of $\xi=0.3$.
\Fref{fig:impactDiskDiameter}(a) shows the corresponding end-to-end distribution
function for $\sigma=a$. The entropic decrease of leaving local void space leads
to a stronger compression of the polymers. This is well seen in the transition
of the main peak from right to left for $p=0.64,0.76,0.89$. Additionally, the
undulations at small end-to-end lengths that represent crumpled configurations
are more pronounced. A further difference is well seen for $p=1$. A narrower
channel favors completely stretched configurations. The intersection between
neighboring void spaces separated by a narrow channel acts as a new pinpoint.
Having a completely occupied lattice leaves only small cavities for the polymer.
For $\xi=0.3$ the length scale of the stiffness is larger than the extension of the
void space. The entropic benefit provided by the larger channels for $\sigma<a$
is not given for $\sigma=a$ and thus stretched configurations contribute for a
major part. 

The extreme case of $\sigma>a$ such that the polymer can no longer
cross between neighboring disks is illustrated in
\fref{fig:impactDiskDiameter}(b). Space is now separated into void-space
clusters. The different contributions hence arise
solely from the different clusters of void space. The fully occupied lattice
finally leaves only small cavities into which the configurations are squeezed. 

\subsection{Leaving the constraint of a fixed pinpoint}\label{sec:leavingConstFixedEnd}

\begin{figure}[t]
  \hspace{-1.3cm}
  \begin{tabular}{c@{\hspace{-2cm}}c}
  \includegraphics{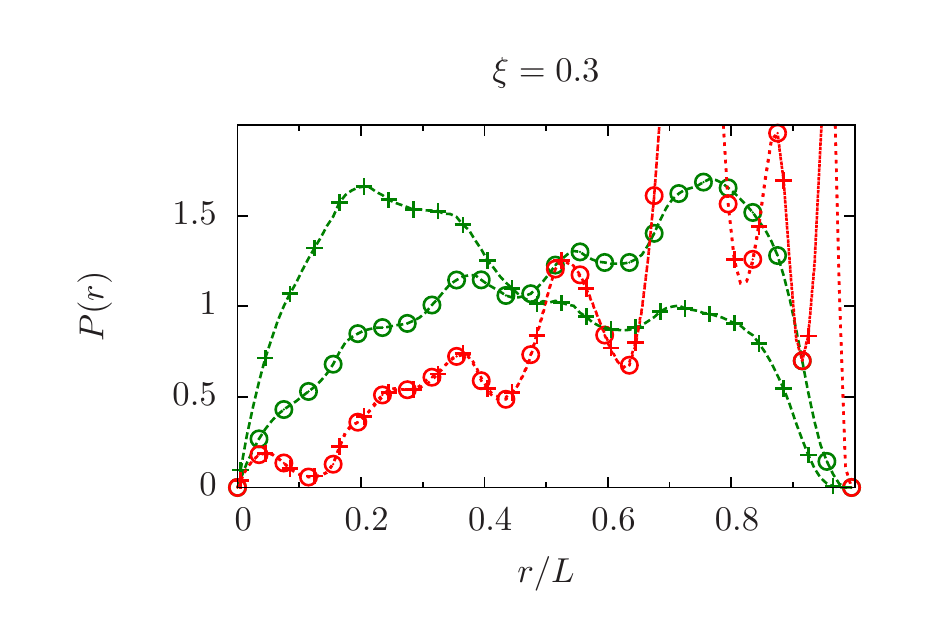}
    &
  \includegraphics{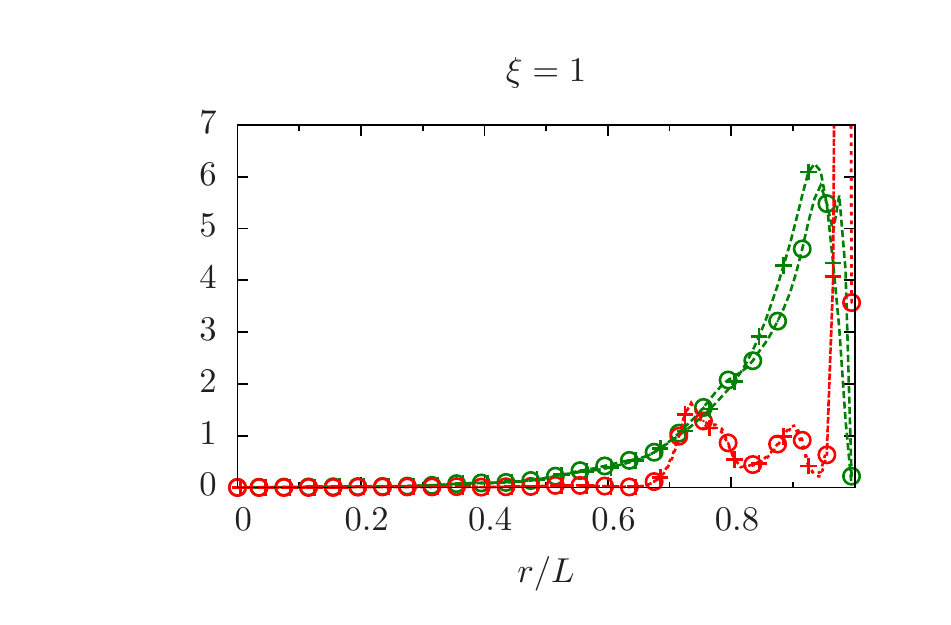}
    \\
    \hspace{1.3cm} (a) & \hspace{1.3cm} (b)
  \end{tabular}
  \caption{(Color online) End-to-end distribution function for $\xi=0.3$ (a)
  and $\xi=1$ (b). The occupation probabilities are: $p=0.89$
  (\hspace{-0.2cm}\protect\includegraphics[scale=1]{greenLongDashed.pdf},
  green), $1.00$
  (\hspace{-0.2cm}\protect\includegraphics[scale=1]{redDashed.pdf},
  red). The data marked by $+$ are for chains that are free to move throughout
  space (no pinpoint) and are done by a multicanonical Monte Carlo simulation.
  The data marked by $\circ$ are for fixed starting point and are obtained with
  the growth method.}
  \label{fig:endToEndN30LatticeNonFixedHigh}
\end{figure}

The discussion so far was subject to the constraint of a fixed pinpoint.  In
this section, we compare results of a non-fixed polymer to the previous case and
discuss the differences that arise.  The data for the non-fixed case originate
from multicanonical Monte Carlo simulations, in which the polymer may move
through space by means of standard rotation and translation updates.  In this
case, we performed longer simulations on each disorder realization and therefore
considered only $300$ of those for the quenched average.
\Fref{fig:endToEndN30LatticeNonFixedHigh} shows results for exemplary
parameters.  It can be seen that for fully occupied lattices the end-to-end
distributions do not differ, as is the case for the free polymer and low
disorder densities.  In the high-density regime, on the other hand, the measured
observables show strong differences especially in the crossover regime of
$\xi=0.3$.  This can be understood considering the following entropic and
energetic arguments.  Other than the fully occupied or the low-density case,
high disorder densities produce small void spaces of different sizes which are
entropically more favorable than the alternative channels.  In the case of
non-fixed constraints, the polymers are able to move to those small spaces and
thus they contribute stronger as long as the energetic cost for bending is not
too high.  The results can be seen in \fref{fig:endToEndN30LatticeNonFixedHigh}
where the end-to-end distribution shows deviations from the case of a fixed
pinpoint in the crossover regime. This effect is less pronounced  for large
persistence lengths, since possible gains in entropy are dominated by the cost
of bending energy.

\section{Conclusions}\label{sec:conclusions}

We analyzed in detail the behavior of a polymer in a potential consisting of
hard disks distributed on the sites of a square lattice. We found that the
polymer, depending on the ratio of persistence length and void space extension,
either crumples up (small $\xi$) or straightens (large $\xi$) for increasing
density of the potential. This is consistent with the results that, e.g., Cifra
\cite{Cifra2012} recently found. Besides, the periodic structure of the lattice is
reflected in the distribution functions of the polymer. 

Furthermore, we found that the distributions---in the case of pinning the polymer
at one end---strongly reflect the local cluster structure of the disorder.
Leaving the constraint of pinning lets the polymer escape local cavities and
gain entropy in larger void-space clusters. The corresponding
distributions for pinned and non-pinned polymers differ considerably.  

Finally, we checked the applicability of an off-lattice growth algorithm to the
problem of a semiflexible polymer exposed to high-density disorder in the form
of steric hindrance. By employing two conceptually completely different
algorithms to the problem---the off-lattice growth algorithm and the
multicanonical Monte Carlo method---we corroborated that the tested method is
well usable. For a combination of large occupation and long persistence length
the growth method is even performing better. We want to emphasize the ability of
the growth algorithm to provide distributions for all chain lengths up to the
desired degree of polymerization within one simulation. Equipped with this
finding, a challenging next step is to investigate the behavior of semiflexible
polymers in hard-disk fluid disorder.

\section*{Acknowledgements}

We thank Klaus Kroy for inspirations to the topic of this work. Furthermore we
thank Sebastian Sturm, Niklas Fricke, and Viktoria Blavatska for fruitful
discussion. In addition we are grateful for support from the Leipzig Graduate
School of Excellence GSC185 ''BuildMoNa'' and the SFB/TRR102 (project B04) as
well as from FOR877 under grant No.\ JA483/29-1 and the Deutsch-Franz\"osische
Hochschule (DFH-UFA) under grant No.\ CDFA-02-07. JZ is grateful for funding by
the European Union and the Free State of Saxony.

\section*{References}
\bibliographystyle{apsrev4-1.bst}

%
%
%
\end{document}